\documentstyle[12pt,epsfig,picinpar]{article}
\begin{document}
 
\title {CHARGINO MASS AND ~$R_b$ ~ANOMALY.
\thanks{Supported in part by the Polish Committee for Scientific
        Research and European Union Under contract CHRX-CT92-0004.}}
\author{Piotr H. Chankowski \\
Institute of Theoretical Physics, Warsaw University\\
ul. Ho\.za 69, 00--681 Warsaw, Poland.\\
\\
Stefan Pokorski 
\thanks{On leave of absence from 
Institute of Theoretical Physics, Warsaw University}
\\
Max--Planck--Institute f\"ur Physik\\
Werner -- Heisenberg -- Institute \\
F\"ohringer Ring 6,
80805 Munich, Germany
}
 
\maketitle
 
\vspace{-12cm}
\begin{flushright}
{\bf IFT-96/6} \\
\end{flushright}
\vspace{12cm}
  
\begin{abstract}
We re-egzamine the possible magnitude of the supersymmetric contribution 
to ~$R_b$ ~with imposed all available phenomenological constraints and 
demanding good quality of the global fit to the precision electroweak data.
For low ~$\tan\beta$ ~we find a new region of the parameter space, with ~
$M_2\approx |\mu|$ ~and ~$\mu<0$ ~where ~$R_b$ ~remains large, ~$\sim0.2180$ ~
even for the lighter chargino as heavy as ~$90-100$ GeV. ~It is an interesting 
mixture of the up-higgsino and gaugino.  The r\^ole of 
various phenomenological constraints is discussed in analitic form
and importance of small but non-negligible left-right  mixing in the stop 
sector is emphasized in this context. The large ~$\tan\beta$ ~option for
enhancement of ~$R_b$ ~is also reviewed. The available data do not rule 
out this scenario.
\end{abstract}

\newpage

{\bf 1. INTRODUCTION.}
\vskip 0.3cm

A considerable excitement has recently been inspired by 
the ~$R_b$ ~and ~$R_c$ ~anomaly 
\cite{RBC_NS,RBC_S,LANER,KANE,MY_MSSM,JA_BRU,SHIF,KAWE,ELONA}. 
The succesful tests of the Standard Model (SM), to a per mille
level \cite{EWWG}, are challenged by the measurements of the 
partial widths of the ~$Z^0$ ~decays
into ~$\overline bb$ ~and ~$\overline cc$ ~quarks which disagree with the 
SM predictions for ~$m_t=180(170)$ GeV ~at the level of 3.7(3.5) and 
2.5(2.5) standard deviations, respectively \cite{BRUSS}. 
If both results are confirmed, the SM and its simplest
supersymmetric extension, the Minimal Supersymmetric Standard Model
(MSSM), are ruled out. However, since
the ~$R_b$ ~anomaly is statistically more significant, it is also of 
interest to discuss the possibility of explaining only the larger than 
in the SM value of ~$R_b$. ~
Even if ~$R_c$ ~is fixed to its SM prediction, ~$R_c=0.172$ ~the then 
measured value of ~$R_b=0.2206\pm0.0016$ is still 3 standard deviations
away from its SM value. 
The issue has been addressed in particular in the framework of the 
MSSM \cite{RBC_S,LANER,KANE,MY_MSSM,JA_BRU,SHIF,KAWE,ELONA}. 
It is well known already for some 
time that in the MSSM there are new contributions to the ~$Z^0\overline bb$ ~
vertex which can significantly enhance the value of ~$R_b$ ~(but not ~$R_c$) ~ 
if some superpartners are sufficiently light 
\cite{BF,RBC_S,KANE,MY_MSSM,JA_BRU,KAWE,ELONA}. More specifically, 
for low (large) ~$\tan\beta$ ~the dominant contributions are chargino--stop
($CP-$odd Higgs boson and chargino--stop) loops. 

Any improvement in  ~$R_b$ ~must not destroy the perfect agreement of the SM 
with the other precision LEP measurements and must be consistent with 
several other experimental constraints (which will be listed later on).
It is, therefore, important to discuss the changes in ~$R_b$ ~
in the context of global fits to the electroweak data (and with all
additional constraints included).
Such fits in the effective low energy MSSM (unconstrained by any GUT 
assumptions about the pattern of soft supersymmetry breaking scalar masses)
have shown that it is realistic to obtain the values of ~$R_b$ ~up to ~
$R_b=0.2180 (0.2190)$ ~for small (large) ~$\tan\beta$ ~values
\cite{MY_MSSM}. Although still away 
by ~$1.5\sigma$ ~($1.0\sigma)$ ~even from the central experimental
value obtained with ~$R_c$ ~fixed to its SM value, those results 
provide an interesting improvement over the SM prediction ~
$R_b=0.2158 ~(0.2160)$ ~for ~$m_t=180 ~(170)$ GeV. ~At the same time the 
overall best ~$\chi^2$ ~is smaller than in the SM fits by  ~
$\Delta\chi^2\approx4$ (5 for the fit with ~
$R_c$ ~fixed to the SM value) ~(for ~$m_t=170$ GeV).
Also, the fitted value of ~$\alpha_s(M_Z)$ ~is modified by ~
$\Delta\alpha_s(M_Z)\approx-4\delta R_b$, ~i.e. lower than in the SM fits 
which give ~$\alpha_s(M_Z)=0.123\pm0.005$ ~\cite{ELLIS,LANER,MY_SM,EWWG}.
This may look 
desirable \cite{SHIF} in view of the apparent hint for some discrepancy
between the value of  ~$\alpha_s(M_Z)$ ~obtained from the SM fits to the
precision electroweak data and the values ~$\alpha_s(M_Z)=0.112\pm0.005$ ~
and ~$\alpha_s(M_Z)=0.112\pm0.007$ ~obtained from the deep inelastic scattering
\cite{DIS} and lattice calculations \cite{LATT}, respectively
\footnote{The overall average gives ~$\alpha_s(M_Z)=0.117\pm0.005$ ~and
          includes also the results obtained from ~$\tau$ ~decay and jet 
          physics \cite{LOW}}.
Furthermore, increase of ~$R_b$ ~in the MSSM implies some light superpartners,
with masses of the order or even below the electroweak scale ~$M_Z$ ~
 \cite{KANE,MY_MSSM,JA_BRU,LEPREP} !
(It is may be worth noting small differences between various viewpoints:
One is to explain the measured value of ~$R_b$. ~The other is to take it 
as a statistical hint for values of  ~$R_b$ ~somewhat higher then
predicted by the SM. It is reasonable to take the latter and to consider 
an increase in ~$R_b$ ~in the range ~0.2170--0.2180(90) ~as interesting.)

In this paper, encouraged by the importance of the ~$R_b$ ~anomaly for 
experimental search for supersymmetry, we take up this issue once again. 
We clarify certain points of the earlier analysis 
and provide further insight into properties of light superpartners 
predicted by such global fits. Furthemore, we clarify the r\^ole of 
various experimental constraints, including the recent new limits on the
chargino mass from LEP1.5. Our main new result is that
for low ~$\tan\beta$ ~the ~$\chi^2$ ~of the global fit and the
value of ~$R_b$ ~depend very weakly on the chargino mass (for fixed stop
mass) in the range 50$-$100 GeV. ~The ~$R_b$ ~remains at the level 
of ~0.2178 ~for ~$m_{C_1}$ ~up to ~90 GeV ~in the region where ~$M_2\sim |\mu|$ ~
and ~$\mu<0$. ~We also discuss large ~$\tan\beta$ ~case.
\vskip 0.5cm

{\bf 2. LOW ~$\tan\beta$ ~REGION.}
\vskip 0.3cm

Our discussion will be divided into small and large ~$\tan\beta$ ~cases.
To start with, let us, however, recall the basic facts from the global
fits in the MSSM. In order to maintain good agreement of the SM with 
the bulk of the precision data, such as ~$\Delta\rho$, ~$M_W$, ~
$\sin^2\theta^{eff}_{lept}$, ... ~we must avoid new sources of the 
custodial ~$SU_V(2)$ ~symmetry breaking in the left currents. 
In the MSSM, this is assured when the left squarks of the third
generation (and all left sleptons) are sufficiently heavy
\cite{MY_DR,MY_MSSM}, say, $>{\cal O}(500~{\rm GeV})$
\footnote{The actual lower limits on left squarks and sleptons depend  on ~
          $\tan\beta$ ~value \cite{MY_DR,MY_MSSM}.}.  
At the same time, an increase in ~$R_b$ ~requires a light right stop.
So, one needs a hierarchy
\footnote{Such a hierarchy is very natural in models where soft scalar
          mass terms have their origin at the GUT scale. Even with universal 
          initial squark mass values ~$m^2_0$, ~the renormalization 
          group evolution with large top quark Yukawa coupling 
          gives the hierarchy (\ref{eqn:hier}) provided ~$m_0^2\gg M_2^2$}:
\begin{eqnarray}
M_{\tilde t_L} >> M_{\tilde t_R}  ~~~{\rm or} ~~~
M_{\tilde t_1} >> M_{\tilde t_2}
\label{eqn:hier}
\end{eqnarray}
(in our notation ~$\tilde t_2$ ~denotes the lighter stop) 
with small left-right mixing.

The second important fact to recall is the pattern of the chargino sector
(masses and mixings). The chargino mass matrix 
\begin{eqnarray}
{\cal L}_{mass} = -{1\over2} (\chi^+,\chi^-)
\left(\matrix{0 & X^T\cr X & 0}\right) \left(\matrix{\chi^+\cr\chi^-}\right)
+ h.c.
\end{eqnarray}
with
\begin{eqnarray}
X=\left(\matrix{M_{2}&\sqrt2M_W\sin\beta\cr\sqrt2M_W\cos\beta &\mu}\right)
\label{eqn:cmass}
\end{eqnarray}
is diagonalized by two unitary matrices ~$Z_+$ ~and ~$Z_-$: ~
$Z_-^T X Z_+= diag(m_{C_1}, m_{C_2})$ ~with ~$0<m_{C_1}<m_{C_2}$  ~(we 
follow the convention and notation of ref. \cite{ROS}) which determine
the projection of the physical two-component states ~$\lambda^{\pm}_i$ 
($i=1,2$) ~on the gaugino and higgsino two-component weak eigenstates ~
$(-i\psi^+, h^+_2,-i\psi^-,h^-_1)\equiv (\chi^+,\chi^-)$ ~
\begin{eqnarray}
h^+_2 = Z_+^{2i}\lambda^+_i, ~~~h^-_1 = Z_-^{2i}\lambda^-_i
\end{eqnarray}
\begin{eqnarray}
\psi^{\pm} = iZ_{\pm}^{1i}\lambda^{\pm}_i
\end{eqnarray}
with the Dirac charginos defined as
\begin{eqnarray}
C^-_i = \left(\matrix{\lambda^-_i\cr\overline\lambda^+_i}\right)
\end{eqnarray}
It is important to notice that a physical state ~$C^-_i$ ~may contain 
different admixtures
of the "up" and "down" higgsinos ~($h_2$ ~and ~$h_1$ ~respectively). For 
instance, it  may be almost pure up-higgsino 
(in which case ~$|Z_+^{2i}|>>|Z_+^{1i}|$) ~
in its lower two-component spinor ~$\lambda^+_i$ ~and almost pure gaugino ~
($|Z_-^{2i}|<<|Z_-^{1i}|$) ~in its upper two-component spinor ~$\lambda^-_i$. ~
For small values of ~$\tan\beta$ ~the pattern of the gaugino-higgsino mixing 
in the physical states is
not just determined by the ratio ~$r\equiv M_2/|\mu|$ ~but also crucially 
depends on the sign of ~$\mu$. ~In addition, for fixed ~$M_2$ ~and ~$\mu$ ~
also the pattern of the physical masses depends on the sign of ~$\mu$ ~
in a crucial way as shown in Fig.1. ~
For better qualitative understanding of those points it is instructive to
consider the chargino masses and mixing for ~$r=1$. ~For both signs 
of the ~$\mu$ ~parameter and any value of ~$\tan\beta$ ~the
chargino mass matrix (\ref{eqn:cmass}) can be easily transformed into 
symmetric form which is diagonalized by an unitary (orthogonal 
in our case of real ~$\mu$ ~and ~$M_2$) ~matrix i.e. ~
$Z_+^{\prime}=Z_-^{\prime}$. ~With both eigenvalues positive and
ordered, ~$m_{C_1}<m_{C_2}$, (we use the fact that ~$\tan\beta>1$), 
the complete diagonalizing matrices read:

\noindent for ~$\mu>0$ ~
\begin{eqnarray}
Z_-=\left(\matrix{-c_{\theta} & s_{\theta} 
               \cr s_{\theta} & c_{\theta}}\right) ~~~
Z_+=\left(\matrix{-s_{\theta} & c_{\theta} 
               \cr c_{\theta} & s_{\theta}}\right) ~~~
\label{eqn:zz_mpos}
\end{eqnarray}

\noindent for ~$\mu<0$ ~
\begin{eqnarray}
Z_-=\left(\matrix{s_{\theta} & -c_{\theta} 
              \cr c_{\theta} &  s_{\theta}}\right) ~~~
Z_+=\left(\matrix{c_{\theta} & -s_{\theta} 
             \cr -s_{\theta} & -c_{\theta}}\right) ~~~
\label{eqn:zz_mneg}
\end{eqnarray}
where $s_{\theta}\equiv\sin\theta$, ~
$c_{\theta}\equiv\cos\theta$. ~For the angle ~
$\theta$ ~we get

\noindent for ~$\mu>0$ ~
\begin{eqnarray}
\tan2\theta = - {2\mu\over\sqrt2M_W(\sin\beta-\cos\beta)} <0 
\end{eqnarray}

\noindent for ~$\mu<0$ ~
\begin{eqnarray}
\tan2\theta = - {2\mu\over\sqrt2M_W(\sin\beta+\cos\beta)} >0 
\end{eqnarray}

We see that for low values of ~$\tan\beta$ ~the two cases are very 
different. For ~$\mu>0$ ~ we get ~$\theta\sim\pi/4$ ~and ~
$c_{\theta}\sim s_{\theta}\sim1/\sqrt2$. ~
Thus all two-component 
states are mixed. Moreover, the mass eigenstates are 
\begin{eqnarray}
m_{C_{1,2}} = \mu\mp{1\over\sqrt2} M_W(\sin\beta+\cos\beta) 
            + {1\over4}(1-\sin2\beta){M^2_W\over\mu} + ...
\end{eqnarray}
for ~$\mu > M_W$ ~(smaller ~$\mu$ ~values are excluded by experimental
constraints) and charginos are split in mass by ~
$\Delta m_C \sim 2M_W$. ~Pure up-higgsino can
only be obtained for ~$r\gg1$. ~One can then
check that the lighter chargino is higgsino-like in both 
two-component spinors i.e. ~$|Z_{\pm}^{21}| >> |Z_{\pm}^{11}|$ ~and 
the second chargino is heavy.

For ~$\mu<0$ ~and ~$|\mu|\ll M_W$ ~we get ~$\theta\sim0$ ~and ~
$c_{\theta}\sim1$. ~The mass eigenvalues for low ~$\tan\beta$ ~are
\begin{eqnarray}
m_{C_1} = \sqrt2\cos\beta M_W\left(1 
        + {1\over2\cos\beta(\sin\beta + \cos\beta)}{\mu^2\over M_W}+...
        \right)\nonumber\\
m_{C_2} = \sqrt2\sin\beta M_W\left(1 
        + {1\over2\sin\beta(\sin\beta + \cos\beta)}{\mu^2\over M_W}+...
        \right)
\end{eqnarray}
First of all, low values of ~$\mu$ are indeed allowed for ~$m_{C_1}>65$ 
GeV ~ and the heavier chargino is still very light. Moreover, the heavier
chargino is pure up-higgsino (gaugino) in its
lower (upper) two-component spinor. For large values of ~$r$ ~and/or large
values of ~$\tan\beta$ ~the difference between ~$\mu>0$ ~and ~
$\mu<0$ ~disappears.

We can study now the supersymmetric contributions to the ~$Z^0\overline bb$ ~
vertex. In the low ~$\tan\beta$ ~region there are 
two types of  relevant diagrams:
with stop coupled to ~$Z^0$ ~and with charginos coupled to ~$Z^0$. ~Their 
(renormalized) contributions to the vector and axial-vector formfactors of 
the ~$Z^0\overline bb$ ~vertex ~$F_V$ ~and ~$F_A$ ~
(${\cal L}^{eff}_{Z^0\overline bb}= \overline\psi_b\gamma^{\mu}
(F_V - \gamma^5 F_A)\psi_b Z^0_{\mu})$ ~are 
(in the limit ~$m_b=0$) ~given by (the sum is over the two stop and two
chargino mass eigenstates):

\begin{eqnarray}
\delta F^{(\tilde t)}_V={e\over4s_Wc_W}\sum_{n,m,l}V_{Z\tilde t\tilde t}^{l,n}
\left(L_{b\tilde tC}^{l,m*} L_{b\tilde tC}^{n,m} 
+ R_{b\tilde tC}^{l,m*} R_{b\tilde tC}^{n,m}\right)\nonumber\\
\times f_{ssf}(M^2_Z;M_{\tilde t_n},m_{C_m},M_{\tilde t_l})\phantom{aaa}
\label{eqn:fv_t}
\end{eqnarray}
\begin{eqnarray}
\delta F^{(\tilde t)}_A
={e\over4s_Wc_W}\sum_{n,m,l}V_{Z\tilde t\tilde t}^{l,n}
\left(L_{b\tilde tC}^{l,m*} L_{b\tilde tC}^{n,m} 
- R_{b\tilde tC}^{l,m*} R_{b\tilde tC}^{n,m}\right)\nonumber\\
\times f_{ssf}(M^2_Z;M_{\tilde t_n},m_{C_m},M_{\tilde t_l})\phantom{aaa}
\label{eqn:fa_t}
\end{eqnarray}

\begin{eqnarray}
\delta F^{(C)}_V={e\over4s_Wc_W}\sum_{n,m,l} \left(
R_{ZCC}^{m,l} L_{b\tilde tC}^{n,l*} L_{b\tilde tC}^{n,m}~
+L_{ZCC}^{m,l} R_{b\tilde tC}^{n,l*} R_{b\tilde tC}^{n,m}\right)\nonumber\\
\times m_{C_m}m_{C_l}~c_0(m_{C_m},M_{\tilde t_n},m_{C_l})
\phantom{aaaa}\\
-{e\over4s_Wc_W}\sum_{n,m,l} \left(
L_{ZCC}^{m,l} L_{b\tilde tC}^{n,l*} L_{b\tilde tC}^{n,m} 
+R_{ZCC}^{m,l} R_{b\tilde tC}^{n,l*} R_{b\tilde tC}^{n,m}\right)\nonumber\\
\times f_{ffs}(M^2_Z,m_{C_m},M_{\tilde t_n},m_{C_l})\nonumber
\phantom{aaaaa}
\label{eqn:fv_c}
\end{eqnarray}
\begin{eqnarray}
\delta F^{(C)}_A={e\over4s_Wc_W}\sum_{n,m,l} \left(
R_{ZCC}^{m,l} L_{b\tilde tC}^{n,l*} L_{b\tilde tC}^{n,m}~
-L_{ZCC}^{m,l} R_{b\tilde tC}^{n,l*} R_{b\tilde tC}^{n,m}\right)\nonumber\\
\times m_{C_m}m_{C_l}~c_0(m_{C_m},M_{\tilde t_n},m_{C_l})
\phantom{aaaa}\\
-{e\over4s_Wc_W}\sum_{n,m,l} \left(
L_{ZCC}^{m,l} L_{b\tilde tC}^{n,l*} L_{b\tilde tC}^{n,m} 
-R_{ZCC}^{m,l} R_{b\tilde tC}^{n,l*} R_{b\tilde tC}^{n,m}\right)\nonumber\\
\times f_{ffs}(M^2_Z,m_{C_m},M_{\tilde t_n},m_{C_l})\nonumber
\phantom{aaaaa}
\label{eqn:fa_c}
\end{eqnarray}
where ~$s_W$, ~$c_W$ ~stand for ~$\sin\theta_W$, ~$\cos\theta_W$ ~and
the finite functions ~$f_{ssf}$, ~$f_{ffs}$ ~and ~$c_0$ ~are defined in the 
Appendix B. The couplings read \cite{ROS}:
\begin{eqnarray}
V_{Z\tilde t\tilde t}^{l,n}&=&T^{1l}T^{1n} 
- {4\over3}\sin^2\theta_W\delta^{ln}
\end{eqnarray}
\begin{eqnarray}
L_{ZCC}^{m,l}&=&Z_+^{1m*}Z_+^{1l} + \delta^{ml}\cos2\theta_W,
\nonumber\\
R_{ZCC}^{m,l}&=&Z_-^{1m}Z_-^{1l*} + \delta^{ml}\cos2\theta_W
\end{eqnarray}
\begin{eqnarray}
L_{b\tilde tC}^{n,l}&=&{e\over s_W}T^{1n}Z_+^{1l}
- {e\over\sqrt2s_Wc_W}{m_t\over M_Z\sin\beta}T^{2n}Z_+^{2l},\nonumber\\
R_{b\tilde tC}^{n,l}&=&
- {e\over\sqrt2 s_Wc_W}{m_b\over M_Z\cos\beta}T^{1n}Z_-^{2l*}
\label{eqn:btc}
\end{eqnarray}
The top squark mixing matrix ~$T^{ij}$ ~is defined by:
\begin{eqnarray}
\left(\matrix{\tilde t_L\cr\tilde t_R}\right) =
\left(\matrix{T^{11} & T^{12}\cr T^{21} & T^{22}}\right)
\left(\matrix{\tilde t_1\cr\tilde t_2}\right) =
\left(\matrix{c_t & -s_t\cr s_t & c_t}\right)
\left(\matrix{\tilde t_1\cr\tilde t_2}\right) 
\label{eqn:tij}
\end{eqnarray}
where ~$s_t\equiv\sin\theta_t, ~c_t\equiv\cos\theta_t$. ~
In the case of small ~$\tan\beta$ ~the couplings ~$R_{b\tilde tC}^{n,l}$ ~
are negligible and we have ~$\delta F_V = \delta F_A$. ~
When the lighter stop, ~$\tilde t_2$, ~is dominantly right-handed, as
required for a large ~$b\tilde t_2C$ ~coupling,
its coupling to ~$Z^0$ ~is suppressed (it is proportional to ~
$g~\sin^2\theta_W$). ~Therefore, 
diagrams with stops coupled directly to ~$Z^0$ ~
cannot give any significant enhancement of ~$R_b$. ~Significant contribution
can only come from diagrams in which charginos are coupled to ~$Z^0$. ~
Their actual magnitude depend on the interplay of the couplings in the ~
$C^-_i\tilde t_2 b$ ~vertex and the ~$Z^0C^-_iC^-_j$ ~
vertex. The first one is large only for charginos with large up-higgsino
component, the second - for charginos with large gaugino component in at least
one of its two-component spinors. As we have seen, this combination 
never happens for ~
$\mu>0$. Large ~$R_b$ ~can then only be achieved at the expense of extremly
light ~$C^-_j$ ~and ~$\tilde t_2$, ~either already ruled out by the existing 
mass limits or in conflict with global ~$\chi^2$ ~and/or other constraints 
such as ~$BR(b\rightarrow s\gamma)$,  ~$BR(t\rightarrow new)$.
In addition, for fixed ~$m_{C_1}$ ~and ~$M_{\tilde t_2}$, ~
$R_b$ ~is larger for ~$r>1$ ~i.e. for higgsino-like chargino as the enhancement
of the ~$C^-_1\tilde t_2 b$ ~coupling is more important than of the ~
$Z^0C^-_1C^-_1$ ~coupling.

For ~$\mu<0$ ~the situation is much more favourable. In the range ~
$r\approx1\pm0.5$ ~the second  chargino which for low ~$\tan\beta$, ~is
very close in mass to the lighter one (Fig.1), has large up-higgsino
component and gaugino component. Large couplings 
in both types of vertices of the diagram with charginos coupled to ~
$Z^0$ ~give significant increase in ~$R_b$ ~even for the lighter 
chargino as heavy as ~$80-90$ GeV ~(similar increase in ~$R_b$ ~for ~
$\mu>0$ ~requires ~$m_{C_1}\approx50$ GeV ~and ~
$M_{\tilde t_2}\approx50$ GeV). ~

We now turn our attention to a global fit to the precision data and to the
r\^ole of the following constraints: ~
1) $\Gamma(Z^0\rightarrow\chi^0_1\chi^0_1)<4$ MeV (in addition to the inclusion
of the decay mode ~$Z^0\rightarrow\chi^0_i\chi^0_j$ ~into the total ~$Z^0$ ~
width in the ~$\chi^2$ ~fit), ~
2) $BR(Z^0\rightarrow\chi^0_1\chi^0_2)<10^{-4}$, ~
3) $M_h>60 ~(50)$ GeV ~for small (large) ~$\tan\beta$,  ~
4) $BR(b\rightarrow s\gamma)$ ~in the range ~
$(1.2 - 3.4)\times10^{-4}$ ~
5) $BR(t\rightarrow new) < 45$\% (following ref. \cite{QUIGG})
\footnote{This is a constraint mainly on the decay ~
          $t\rightarrow \tilde t_2 N^0_i$. ~In the large ~$\tan\beta$ case,
          an important decay can also be ~$t\rightarrow bH^+$, ~with ~
          $M_{H^{\pm}}$ ~close to ~$M_W$. ~It is difficult to distinguish this 
          decay mode from the standard one, ~$t\rightarrow bW^+$ ~
          \cite{CPYUAN} and to put an experimental upper bound on 
          this branching ratio.}.  ~
6) Recent exclusion curves in the ~$(m_{N_1}, ~M_{\tilde t_2})$ ~plane from ~
$D0$ ~obtained under the assumption ~
$m_{C_1}>M_{\tilde t_2}$ ~\cite{D0_EXCL}. ~
Since the uncertainty in most of these
constraints is much larger than in the precision data, we perform the ~
$\chi^2$ ~fit to the latter and impose the former as rigid constraints.
Importance of the constraints is illustrated in Figs. 2 and 3 ~and should be 
discussed in more detail. 

The limits ~$\Gamma(Z^0\rightarrow\chi^0_1\chi^0_1)<4$ MeV ~and ~
$BR(Z^0\rightarrow\chi^0_1\chi^0_2)<10^{-4}$ ~put a constraint on the ~
$(M_2,~\mu)$ ~parameter space provided we make an additional assumption about
the values of ~$M_1$ ~(bino mass). In this paper, for a sake of definiteness,
we adopt the GUT relation ~$M_1 = (5/3)\tan^2\theta_W M_2$. ~Then for low ~
$\tan\beta$ ~the two constraints eliminate (approximately) a band in the~
$M_2$, ~$\mu$ ~plane bounded by ~$-50<\mu <100$ GeV. ~ 
This is basically due to the fact that for ~$M_2$ ~and ~$\mu$ ~in this region 
neutralinos are too light and/or have too strong coupling to ~$Z^0$. ~
The values of ~$M_2$ ~and ~$\mu$ ~chosen for Figs. 2,3 (and 7) ~are outside the
forbidden region.
One should remember, however, that the limits on ~$Z^0\rightarrow N^0_1N^0_1$ ~
and ~$Z^0\rightarrow N^0_1N^0_2$ ~are less constraining for larger values of ~
$M_1$, ~$M_1>0.5M_2$ ~(i.e. heavier LSP). Similarly, the bounds on ~
$M_{\tilde t_2}$ ~from the ~$D0$ ~exclusion curves gradually disappear for ~
$M_1>0.5M_2$. ~Finally, the larger the ~$M_1$ ~the better the degeneracy
between the lighter chargino and the lightest neutralino masses.  This is 
important since the new LEP1.5 limit ~$m_{C_1}>65$ GeV ~has been obtained 
under the assuption ~$m_{C_1}-m_{N_1^0}>10$ GeV ~\cite{ALEPH}.
We see that the significance
of various constraints crucially depends on the ratio ~$M_1/M_2$. With the GUT
assumption, our results remain on the conservative side.

The r\^ole of the lower limit on the Higgs boson mass (for a compact formula
for radiatively corrected lighter Higgs boson mass in the limit ~$M_A>>M_Z$ ~
see \cite{HEMPF}) depends on the mass of the heavier stop and  the 
left-right mixing angle. For ~$M_{\tilde t_1} > 500$ GeV ~(as required for
good quality of the global fit) and small mixing angles (necessary for
large ~$R_b$) ~$M_h$ ~is above the experimental limit 
\footnote{Important r\^ole of the experimental lower bound on ~$M_h$ ~
          in ref. \cite{ELONA} in constraining the potential increase of ~
          $R_b$ ~is due to the 
          chosen upper bound ~$M_{\tilde t_1} < 250$ GeV ~which, anyway,  
          looks too low from the point of view of global fit.}
in a large range of the parameter space. Very small and
large left-right mixing angles are, however, ruled out by this constraint.
This is clearly seen in Figs. 2 and 3 ~ where we show the allowed region 
in the ~$(M_{\tilde t_2}, \theta_t)$ ~plane for fixed  ~$M_2$, ~$\mu$ ~and ~
$\tan\beta$. 

The ~$b\rightarrow s\gamma$ ~decay is a very important constraint on the 
parameters space.
In addition to the experimental error in the ~$BR(b\rightarrow s\gamma)$ ~
(which we take at the ~$2\sigma$ ~level) there is large uncertainty in 
the theoretical prediction mainly due to its renormalization scale
dependence \cite{BMMP}. In Figs. 2 and 3 ~we show the significance of the ~
$b\rightarrow s\gamma$ ~constraint in two cases: when the renormalization 
scale ~$Q$ ~ is fixed, ~$Q=m_b=4.7$ GeV ~and with the theoretical 
uncertainty included. In the latter case we consider the result for the ~
$BR(b\rightarrow s\gamma)$ ~as acceptable if any of the theoretical 
values obtained with ~$Q$ ~varying from ~$m_b/2$ ~to ~$2m_b$ ~falls
into the ~$2\sigma$ ~experimental range. Moreover, the results for ~
$BR(b\rightarrow s\gamma)$ ~show weak but nonnegligible dependence 
on the value of ~$\alpha_s(M_Z)$. ~Since the fitted value ~
$\alpha_s(M_Z)$ ~depends on the change in  ~$R_b$, ~in Figs. 2 and 3 ~for 
self-consistency we use the value ~
$\alpha_s(M_Z)=0.123+\delta\alpha_s$ ~where ~
$\delta\alpha_s=-4\delta R_b$ ~and the value ~0.123 ~is obtained from 
the SM fit \cite{ELLIS,LANER,MY_SM,EWWG}. For comparison, in Fig. 2,  we also 
show the regions excluded by this constraint with ~$\alpha_s(M_Z)$ ~
fixed to two different values ~0.114 ~and ~0.135. ~An important 
message from Figs. 2 and 3 ~is the dependence of the ~$b\rightarrow s\gamma$ ~
constraint on the left-right mixing in the stop sector. One should
stress that we keep the mass of the ~$CP-$odd Higgs boson large, ~
$M_A=1$ TeV ~(as needed for large ~$R_b$) ~and in consequence
the charged Higgs boson is also heavy and its exchange  gives 
negligible contribution to the ~$BR(b\rightarrow s\gamma)$. ~
The acceptable values of this branching ratio are obtained from the
sum ~$|{\cal A}^{b\rightarrow s\gamma}_{W} +
{\cal A}^{b\rightarrow s\gamma}_{SUSY}|$ ~
of the SM ~$W^{\pm}$ ~exchange and the supersymmetric
contribution of the ~$\tilde t-C^-$ ~loops which  has to be
of the opposite sign. There are two 
possible solutions: either ~$|{\cal A}^{b\rightarrow s\gamma}_{SUSY}| 
\ll |{\cal A}^{b\rightarrow s\gamma}_{W}|$ ~or ~
$|{\cal A}^{b\rightarrow s\gamma}_{SUSY}| \gg
|{\cal A}^{b\rightarrow s\gamma}_{W}|$. ~
The strong dependence of the supersymmetric contribution to the ~
$b\rightarrow s\gamma$ ~rate on the mixing angle ~$\theta_t$ ~
observed in Fig. 2 and 3 ~can be understood from the general formulae 
given in \cite{BSG}. For ~$r=1$ ~and small angles ~$\theta_t$ ~(interesting
values of ~$\Delta R_b$ ~are obtained only for small mixing angles ~
$\theta_t$; ~large angles are anyway eliminated by the constraint from 
the lighter Higgs boson mass), setting ~$\sin^2\theta_t\approx0$, ~
neglecting the contribution of the heavier stop ~$\tilde t_1$ ~
and using the explicit form of the matrices ~$Z_+$ ~and ~$Z_-$, ~eqs.
(\ref{eqn:zz_mpos},\ref{eqn:zz_mneg}), we get the expression:
\begin{eqnarray}
{\cal A}^{b\rightarrow s\gamma}_{\gamma/gl, ~SUSY} &=& 
-g_t^2 c_{\theta}^2{M^2_W\over m_{C_1}^2}
 f^{(1)}_{\gamma/gl}\left({M^2_{\tilde t_2}\over m^2_{C_1}}\right)
-g_t^2 s_{\theta}^2{M^2_W\over m_{C_2}^2}
 f^{(1)}_{\gamma/gl}\left({M^2_{\tilde t_2}\over m^2_{C_2}}\right)
\nonumber\\
&+&\sin2\theta_t ~s_{\theta}c_{\theta}g_t\left[
{M_W^2\over m_{C_1}^2}
f^{(1)}_{\gamma/gl}\left({M^2_{\tilde t_2}\over m_{C_1}}\right)
+\sigma {M_W^2\over m_{C_2}^2}
f^{(1)}_{\gamma/gl}\left({M^2_{\tilde t_2}\over m_{C_2}}\right)
\right.\nonumber\\
&+&\left.\sigma {M_W\over 2m_{C_1}}
 f^{(3)}_{\gamma/gl}\left({M^2_{\tilde t_2}\over m_{C_1}}\right)
+\sigma {M_W\over 2m_{C_2}}
f^{(3)}_{\gamma/gl}\left({M^2_{\tilde t_2}\over m_{C_2}}\right)\right]
\end{eqnarray}
where ~$\sigma\equiv sign(\mu)$, ~
$g_t\equiv m_t/\sqrt2M_W\sin\beta$ ~and the expressions for the
functions ~$f^{(i)}_{\gamma/gl}(x)$ ~can be found in \cite{BSG}.
Recalling that, for the same values of the arguments, functions ~
$f^{(3)}_{\gamma/gl}(x)$ ~are roughly 5 times larger in absolute values than 
the ~$f^{(1)}_{\gamma/gl}(x)$'s ~(and both are negative) 
it is easy to see that for ~$\mu>0$ ~
${\cal A}^{b\rightarrow s\gamma}_{SUSY}$ ~is small for ~$\theta_t>0$ ~due to
the cancellation of terms of order ~$g^2_t$ ~with those of order ~$g_t$. ~
For small negative ~$\theta_t$ ~both types of terms add up and cancel with
the standard ~$W^{\pm}$ ~contribution making the total amplitude
unacceptably small. Eventually, for large negative angles, the supersymmetric
contribution overcomes the standard one making again the total amplitude
of right magnitude. However this region is excluded by the 
experimental limits on the Higgs boson mass.
For negative ~$\mu$ ~the cancelation between terms of order ~
$g^2_t$ ~and those of order ~$g_t$, ~which makes ~
${\cal A}^{b\rightarrow s\gamma}_{SUSY}$ ~small enough, occurs for 
negative angles ~$\theta_t$. ~For positive angles the supersymmetric 
contribution is large and cancels the standard one making the total amplitude 
too small. 

Thus, there are two mechanisms for obtaining an acceptable value
for ~$BR(b\rightarrow s\gamma)$ ~in the MSSM. One is a cancellation 
between the ~$H^{\pm}$ ~and supersymmetric contributions \cite{ZWIR}.
This mechanism is, however, essentially  in conflict with the simultaneous
increase of ~$R_b$ ~due to large negative contribution of the ~$H^{\pm}$ ~
exchange to ~$R_b$. ~The other mechanism is based on the choice of the 
proper range of the left-right mixing angles.
It is certainly interesting to notice that for small ~$\tan\beta$ ~the 
region of acceptable ~
$BR(b\rightarrow s\gamma)$ ~partly overlaps with the region of large ~$R_b$. ~
Small asymmetry (with respect to ~$\theta_t=0$ ~of the constant ~
$\Delta R_b^{SUSY}$ ~contours seen in Figs. 2 and 3 ~can also be traced back to
the ~$\theta_t$ ~dependence of the couplings ~$L^{n,l}_{b\tilde tC}$ ~eqn.
(\ref{eqn:btc}).

The exclusion curve from the condition ~
$BR(t\rightarrow\tilde t_2N^0_i)<45$\% ~
is also shown in Fig. 2. ~The dependence of ~
$BR(t\rightarrow\tilde t_2N^0_i)$ ~on the left - right mixing
in the stop sector, i.e. on the angle ~$\theta_t$, ~can be understood from 
the formula for the decay width ~$\Gamma(t\rightarrow\tilde t_2N^0_i)$: ~
\begin{eqnarray}
\Gamma(t\rightarrow\tilde t_kN^0_i)&=&{e^2\over s^2_Wc^2_W}{m_t\over64\pi}
\sqrt{1-2(x_k+y_i) + (x_k-y_i)^2}\nonumber\\
&\times&\left[(|c^{ki}_L|^2 + |c^{ki}_R|^2)(1+y_i - x_k) + 
4{\cal R}e(c^{ki*}_Lc^{ki}_R)\sqrt{y_i}\right]
\end{eqnarray}
where ~$x_k\equiv M^2_{\tilde t_k}/m_t^2$, ~$y_i\equiv m^2_{N_i}/m_t^2$ ~
and the couplings read \cite{ROS}:
\begin{eqnarray}
c^{ki}_L&=&\left({1\over3}s_WZ_N^{1i}+c_WZ_N^{2i}\right)T^{1k}
         + {m_t\over\sin\beta M_Z}Z_N^{4i} T^{2k}\nonumber\\
c^{ki}_R&=&-{4\over3}s_WZ_N^{1i} ~T^{2k}
         + {m_t\over\sin\beta M_Z}Z_N^{4i} ~T^{1k}
\end{eqnarray}
and ~$Z^{ji}_N$ ~diagonalizes the neutralino mass matrix \cite{ROS}.
For example, for ~$\mu<0$ ~and ~$r=1$, ~$\Gamma(t\rightarrow new)$ ~is
dominated by ~$\Gamma(t\rightarrow \tilde t_2N^0_2)$ ~because the second
neutralino has larger up-higgsino component. In that case 
\begin{eqnarray}
c^{22}_L\approx  {m_t\over\sin\beta M_Z}Z_N^{42}\cos\theta_t\nonumber\\
c^{22}_R\approx -{m_t\over\sin\beta M_Z}Z_N^{42}\sin\theta_t
\end{eqnarray}
and
\begin{eqnarray}
\Gamma(t\rightarrow\tilde t_2N^0_2)&=&{e^2\over64\pi s^2_Wc^2_W}
{m^3_t\over M^2_Z\sin^2\beta}|Z^{42}_N|^2
\sqrt{1-2(x_2+y_2) + (x_2-y_2)^2}\nonumber\\
&\times&\left[(1+y_2 - x_2) - 
2\sin2\theta_t\sqrt{y_2}\right]
\end{eqnarray}
explaining larger ~$BR(t\rightarrow new)$ ~for negative ~$\theta_t$ ~seen
in Fig. 2.

The ~$D0$ ~exclusion contours \cite{D0_EXCL} in 
the ~$(M_{\tilde t_2}, ~m_{N^0_1})$ ~plane eliminate in Fig. 2 a band ~
$70 ~{\rm GeV}< M_{\tilde t_2}< m_{C_1}$ ~(not shown) and a similar band 
in Fig. 3. This follows from Fig. 10 in ref. \cite{D0_EXCL} (in Fig. 2 ~
$M_2 = - \mu = 58$ GeV, ~$m_{C_1}=85$ GeV ~and ~$m_{N_1}\approx30$ GeV ~
(with the GUT assumption)). Finally, it is worth recalling that 
\begin{eqnarray}
M^2_{\tilde t_2} = M^2_{\tilde t_R} - \theta^2_t M^2_{\tilde t_L}, ~~~
|\theta_t| = \left|{m_tA_t\over M^2_{\tilde t_L}}\right| \ll 1
\end{eqnarray}
where 
\begin{eqnarray}
M^2_{\tilde t_R} = m^2_T + m^2_t + {2\over3}\cos2\beta (M^2_Z-M^2_W)
\end{eqnarray}
($m^2_T$ ~is the soft supersymmetry breaking mass term). If we required ~
$m^2_T>0$ ~then with ~$M_{\tilde t_L}\approx1$ TeV ~we 
get ~$|\theta_t| > 9^o$ ~
for ~$M_{\tilde t_2}\sim{\cal O}(100$ GeV) ~and with ~
$M_{\tilde t_L}\approx1.5$ TeV ~$-$ ~$|\theta_t| > 6^o$. ~The corresponding 
values of ~$A_t$ ~are ~$\sim1$ TeV ~and ~$\sim600$ GeV ~respectively. 
The theoretical status of the condition ~
$m^2_T>0$ ~remains, however, unclear and we do not impose it on our global
fit (with this condition imposed the fit would not have changed too
significantly in particular given the freedom in the choice of ~
$M_{\tilde t_L}$).

The results of a global fit are presented in Figs. 4-6 ~as projections of ~
$\chi^2$ ~as a function of ~$m_{C_1}$ ~for several values of ~$r$ ~and two
different lower bounds for the scan in ~$M_{\tilde t_2}$. ~We also show
the best values of ~$R_b$ ~obtained with the restriction 
~$\chi^2 < \chi^2_{min} +1$ ~where ~$\chi^2_{min}$ ~is the minimum of the ~
$\chi^2$  ~for fixed ~$m_{C_1}$. ~The fitting procedure
is the same as in ref. \cite{MY_MSSM}. In particular the fitted parameters 
include ~$m_t$, ~$\tan\beta$, ~$M_2$, ~$\mu$, ~$\alpha_s(M_z)$, ~
$M_{\tilde t_2}$ ~and ~$A_t$. ~The values of the parameters
$M_{\tilde t_L}$, ~$M_{A^0}$, ~$M_{\tilde l_L}$ ~provided 
large enough, are not relevant for the quality of the fit and we have fixed 
them at ~1 TeV. ~Similarly, the parameters like ~$M_{\tilde l_R}$, ~
$M_{\tilde b_R}$, ~masses 
and mixings of the two first generations of squarks to which the fit is 
not sensitive have been fixed at ~1 TeV ~(in the large ~$\tan\beta$ ~case 
the best fit is obtained
for ~$M_{\tilde b_R}=130$ GeV ~due to additional contribution of ~
$N^0_i-\tilde b_R$ ~loops to ~$R_b$).

The value of ~$m_t=170$ GeV  ~chosen for the plots is close to the 
best value obtained from the fit. Larger values 
of ~$m_t$ ~give worse ~$\chi^2$ ~and this is 
a reflection of the well known from the SM fits correlation between the 
Higgs boson and the top quark masses \cite{EWWG,MY_SM}. Here the ~$M_h$ ~
is constrained by supersymmetry and cannot follow the increase of ~$m_t$. ~
In our fit ~$\tan\beta$ ~is bounded so that top quark Yukawa coupling 
remains perturbative up to the GUT scale (we take ~$\tan\beta\geq1.4$ ~for ~
$m_t=170$ GeV ~and ~$\tan\beta\geq1.6$ ~for ~$m_t=180$ GeV). ~We impose this 
theoretical constraint to remain on the conservative side, as lower values 
of ~$\tan\beta$ ~(for the same ~$m_t$) ~give larger ~$R_b$. ~For a given ~
$m_t$ ~the best fit is obtained for the lowest value of ~$\tan\beta$ ~(for
very large  ~$\tan\beta$ ~see next section). One may also wish to impose
the constraint ~$M_2>50$ GeV ~which follows from the lower experimental
bound on the gluino mass ~$m_{\tilde g} > 150$ GeV ~and the assumed GUT
relation for the gaugino masses. This constraint does not change the best
values of ~$R_b$ ~and ~$\chi^2$ ~(but it eliminates e.g. large part of our
curve with ~$r=0.5$, ~$\mu<0$. ~

All discussed earlier constraints are included in those results.
The fit reveals several interesting facts. For both signs of ~$\mu$, ~the
value of ~$\chi^2$ ~eventually increases towards low values of ~$m_{C_1}$. ~
The reason for this behavior of ~$\chi^2$ ~is the contribution of the 
neutralinos to ~$\Gamma_Z$ ~ which becomes too large and correspondingly ~
$\sigma_{hadr}$ - too small (at this point we should again remember about our 
assumption ~$M_1 = (5/3)\tan^2\theta_W M_2$). ~This effect is much stronger 
than the decrease of ~$\Gamma_Z$ ~due to the chargino contribution 
to the ~$Z^0$ ~boson wave function renormalization \cite{BFC}. 
So, the increase of ~$R_b$ ~is bounded not only by the lower experimental
limits on ~$m_{C_1}$, ~$M_{\tilde t_2}$ ~but also by the quality of the fit
to the precision data. Still, for ~$\mu>0$, ~the best fit is obtained for ~
$m_{C_1}\sim50$ GeV, ~well below the new experimental limit from LEP1.5, ~
$m_{C_1}>65$ GeV ~ \cite{ALEPH}. 
This limit is valid for ~$m_{C_1} - m_{N_1} > 10$ GeV. ~In principle, the 
degeneracy of the chargino and neutralino masses can be better than ~10 GeV. ~
This occurs for ~$r>10$ ~with ~$M_1\approx0.5M_2$ ~
(note however, that supersymmetric contribution
to ~$R_b$ ~decreases for very large values of ~$r$) ~or for any value of ~$r$ ~
for sufficiently large ~$M_1$. ~This is an obvious possibility which we 
do not discuss further. 
On the other hand, for ~$m_{C_1}=65$ GeV ~the maximal reachable value of ~
$R_b$ ~is only ~$R_b=0.2173$. 

For ~$\mu<0$ ~due to the mechanism explained earlier, ~$\chi^2$ ~($R_b$) ~is
small (large) for much larger values of ~$m_{C_1}$. ~
In fact the new LEP1.5 limit is totally irrelevant in this case.
A chargino with mass ~$70-90$ GeV ~and with the composition described by ~
$M_2\approx |\mu|$ ~remains an interesting potential possibility. 
The dependence of ~$R_b$ ~on the stop mass can be inferred from
comparison of Fig. 4 and 5. We see that, even for ~
$M_{\tilde t_2}= m_{C_1}=90$ GeV, ~$R_b>0.2175$. ~Moreover, a significant
enhancement in ~$R_b$ ~is consistent with both configurations: ~
$M_{\tilde t_2}> m_{C_1}$ ~and ~$M_{\tilde t_2} < m_{C_1}$. ~In Fig. 1 ~
we show the cross sections for chargino production as ~$\sqrt{s}=190$ GeV. ~
The corresponding cross sections 
for the production of the pair ~$N^0_1N^0_2$ ~ are also shown. 
\vskip 0.5cm

{\bf 3. LARGE ~$\tan\beta$.}
\vskip 0.3cm

Significant enhacement of ~$R_b$ ~is also possible for large ~$\tan\beta$ ~
values, ~$\tan\beta\approx m_t/m_b$. ~In this case, in 
addition to the stop - 
chargino contribution there can be even larger positive contribution from 
the ~$h^0$, ~$H^0$ ~and ~$A^0$ ~exchanges in the loops, provided those
particles are sufficiently light (in this range of ~$\tan\beta$ ~
$M_h\approx M_A$) ~and non-negligible sbottom-neutralino loop 
contributions. Recently it has been argued \cite{KAWE} that large ~
$\tan\beta$ ~option of an enhancement in  ~$R_b$ ~is already ruled out by
the constraints from ~$b\rightarrow c\tau\overline\nu_{\tau}$ ~and the 
absence of ~$4b$ ~ events in the ~$Z^0$ ~decay which should have been present
as a signature for the
bremstrahlung production of a light pseudoscalar Higgs boson ~$A^0$ ~at
LEP1: ~$Z^0\rightarrow\overline bbA^0\rightarrow\overline bb\overline bb$. ~
It has been claimed that the combination of the two constraints rules out 
the region in the ~$(M_A, ~\tan\beta)$ ~plane which can give substantial 
increase of ~$R_b$ ~(see Fig. 2 in ref. \cite{KAWE}). It seems, however, 
that this conclusion cannot be maintained after proper account of 
experimental efficiencies in the search for the process ~
$Z^0\rightarrow\overline bb\overline bb$. ~The decay ~
$b\rightarrow c\tau\overline\nu_{\tau}$ ~with ~$BR=2.69\pm0.44$ \% ~
indeed puts an upper
bound ~$\tan\beta< 0.52 M_{H^{\pm}}/(1GeV)$ ~\cite{BCTAU} 
which translates into a bound on ~$M_A$ ~as a function of ~$\tan\beta$ ~
which is given in ref. \cite{KAWE}. For instance, for ~$M_A=55 ~(65)$ 
GeV, allowed ~$\tan\beta<63 ~(68)$. ~
However, with proper account of the experimental efficiency in search for ~
$4b$ ~events and with ~$10^7$ ~events from LEP1, 
one cannot expect to get for, say, ~$\tan\beta=60$, ~a lower limit on ~
$M_A$ ~from this production mechanism better than about ~$40$ GeV ~
(since the four ~$b$ ~ tagging procedure is not 100\% efficient, 
the main source of the QCD background are ~$\overline bb gg$ ~ and ~
$\overline bb\overline qq$ ~final states) 
\footnote{S. Simion and P. Janot, private communication. 
          Such an experimental analysis has 
          not yet been performed. We thank J. Kalinowski, M. Krawczyk,
          M. Carena and C. Wagner for several discussions of this point.}. ~
This is well below the reported limit 
of ~$55$ GeV ~based on the analysis of the process ~
$Z^0\rightarrow A^0h^0$ ~\cite{AMASS_LIM}. Thus, an enhancement of ~$R_b$ ~in 
the large ~$\tan\beta$ ~region still remains an open possibility and we briefly
summarize the relevant results.

The discussion of large ~$\tan\beta$ ~case is relatively simpler as
most of the results is symmetric under the simultaneous change of sign of ~
$\mu$ ~and ~$\theta_t$. ~The contribution to the ~$Z^0\overline bb$ ~
vertex from charginos-stop loops to the vector and axial-vector 
formfactors are (in the limit ~$m_b=0$) ~given by  
eqs. (\ref{eqn:fv_t}-\ref{eqn:fa_c}). 
Compact expressions for the remaining  contributions are
collected in Appendix A.

In the large ~$\tan\beta$ ~case, for both signs of ~$\mu$ ~the 
chargino composition 
is the same in the up and down Weyl spinors and is a monotonic function of the
ratio ~$r$. ~For enhancement in ~$R_b$ ~the ~$b\tilde t_2C$ ~coupling is more 
important than the ~$Z^0CC$ ~coupling, so higgsino-like chargino ~$(r\gg1)$ ~is
more favourable. The chargino masses are then given by
\begin{eqnarray}
m_{C_1}&\approx&|\mu| ~\left(1 - {M^2_W\over M_2^2} + ...\right)\nonumber\\
m_{C_2}&\approx& M_2 ~\left(1 + {M^2_W\over M^2_2} + ...\right)
\end{eqnarray}
The diagonalizing matrices have the form
\begin{eqnarray}
Z_{\pm}=\left(\matrix{-\sin\theta_{\pm} &\cos\theta_{\pm}\cr
                       \cos\theta_{\pm} &\sin\theta_{\pm}}\right)
\end{eqnarray}
with
\begin{eqnarray}
\tan2\theta_- = {2\sqrt2M_W\mu\over M^2_2 + 2M^2_W -\mu^2}, ~~~
\tan2\theta_+ = {2\sqrt2M_WM_2\over M^2_2 - 2M^2_W -\mu^2}
\end{eqnarray}
so that indeed for ~$M_2\gg\mu, M_W$ ~both angles are close to zero. Notice
however, that ~$\theta_+$ ~(whose smallness is  important for the coupling ~
$b\tilde t_2 C$) ~approaches zero much slower than ~$\theta_-$ ~as ~$r$ ~
is increased.
The r\^ole of the experimental constraints (listed in the previous section)
on the parameter space is illustrated
in Fig. 7 ~The regions ruled out by the ~
$BR(b\rightarrow s\gamma)$ ~are easy to understand.
Neglecting the contribution of the heavier chargino and stop and restricting
to small angles ~$\theta_t$ ~in the general formula of \cite{BSG} we get for 
the ~$b\rightarrow s\gamma$ ~amplitude the following expression
\begin{eqnarray}
{\cal A}^{b\rightarrow s\gamma}_{\gamma /gl, ~SUSY}\approx
-g^2_t \cos^2\theta_+ {M^2_W\over m_{C_1}^2} 
f^{(1)}_{\gamma /gl}\left({M^2_{\tilde t_2}\over m^2_{C_1}}\right)
\phantom{aaaaaaaaaaaaaa}\nonumber\\
+\sin2\theta_t\left({\tan\beta\over2\sqrt2}\right) g_t \cos\theta_+
\cos\theta_- {M_W\over m_{C_1}} 
f^{(3)}_{\gamma /gl}\left({M^2_{\tilde t_2}\over m^2_{C_1}}\right)\\
+ {1\over2}{m^2_t\over M^2_{H^{\pm}}}~
f^{(2)}_{\gamma /gl}\left({m^2_t\over M^2_{H^{\pm}}}\right)
\phantom{aaaaaaaaaaaaaa}\nonumber
\label{eqn:bsg_l}
\end{eqnarray}
With very light ~$A^0$, ~the charged Higgs boson contribution is large and 
adds up (with the same sign) to the standard ~$W^\pm$-top exchange amplitude.
Thus we need a cancellation from the supersymmetric part of 
eq. (\ref{eqn:bsg_l}).
Due to the presence of large ~$\tan\beta$ ~factor, the second term
is strongly dominant. Therefore, for ~$\theta_t \leq 0$ ~supersymmetric
contribution to the ~$b\rightarrow s\gamma$ ~amplitude is of the
opposite sign to the standard ~contribution. Moreover, for
angles ~$\theta_t$ ~not very close to zero, the absolute value of the 
supersymmetric contribution exceeds the standard one. Thus, for a given 
chargino mass there are only two very narrow bands seen in Fig. 7 in the 
plane ~$(\theta_t, ~M_{\tilde t_2})$ ~where the total amplitude is acceptable:
in the right one the total amplitude is of the same sign as the standard ~
$W^{\pm}-$top ~and charged Higgs boson amplitudes and  in the left band
it is of the opposite sign. From the Figure 7 
it is however clear that for large ~$\tan\beta$, ~the ~
$b\rightarrow s\gamma$ ~rate does not constrain the value of ~$R_b$ ~at all 
\cite{MY_MSSM}.

The lower experimental bound on the lightest Higgs boson mass in the large ~
$\tan\beta$ ~scenario is ~$\sim40$ GeV. Since in the MSSM ~
$M_{A^0}\approx M_{h^0}$, ~our results are not constrained by this bound.

In the parameter space which gives enhancement in ~$R_b$, ~also the decays ~
$t\rightarrow new$ ~are enhanced. In addition 
to ~$t\rightarrow \tilde t_2N^0_i$, ~important is also the decay ~
$t\rightarrow bH^+$. ~For instance, for ~$m_t=170$ GeV, ~$\tan\beta=50$, ~
$m_{C_1}=65$ GeV ~and ~$M_A=55 ~(65)$ GeV ~we get ~$BR(t\rightarrow bH^+)$ ~
ranging from 37(34)\% ~up ~to 49(46)\% ~depending on the stop 
sector parameters. ~One should remember, however, that the decay ~
$t\rightarrow bH^+$ ~is not easily
distinguishable from the standard one, ~$t\rightarrow bW^+$, ~for ~$M_{H^+}$ ~
close to ~$M_W$ ~\cite{CPYUAN} and we do not impose this constraint in
the global fit.

In Figs. 8 and 9 ~we present the results of a global fit in the 
large ~$\tan\beta$ ~
case, together with the corresponding values of ~$R_b$. ~Two important features
of the global ~$\chi^2$ ~is the strong decrease in the quality of the fit for ~
$m_t=180$ GeV (compared to the best fit for ~$m_t=170$ GeV) ~and for light 
charginos (below 60 GeV). ~ Those effects are stronger than similar effects
for low ~$\tan\beta$.

The values of ~$R_b$ ~are almost insensitive to 
the value of ~$M_{\tilde t_2}$ ~in the range ~$50 - 100$ GeV ~and
show the expected dependence on ~$M_A$ ~and ~$m_{C_1}$ ~
with the maximal values for very light
charginos. It is, therefore, worth recalling that the new
limit ~$m_{C_1}>65$ GeV ~is based
on the assupmtion ~$m_{C_1}-m_{N_1}>10$ GeV. As for low ~$\tan\beta$, better
degeneracy of the two masses can be achieved for ~$r>10$ ~and/or
for ~$M_1 > 0.5M_2$. ~In this case also the quality of the fit with charginos
below ~65 GeV ~is improved as heavier neutralinos contribute 
less to ~$\Gamma_Z$. ~It is also remarkable, that due to the combined
effect of neutral Higgs exchange and the chargino-stop together with the 
neutralino - sbottom contributions, the ~$R_b$ ~remains greater than ~0.2175 ~
for masses well above the present experimental limits. E.g., for  ~
~$m_{C_1}\approx M_{\tilde t_2}\approx M_A\approx70$ GeV, ~$R_b=0.2178$. ~
\vskip 0.5cm

{\bf 4. CONCLUSIONS.}
\vskip 0.3cm

In this paper we re-examined the possible magnitude of the supersymmetric 
contributions to ~$R_b$, ~with imposed all available phenomenological 
constraints and demanding good quality of the global fit to the precision 
electroweak data. For low ~$\tan\beta$ ~we have found a new region of the 
parameter space, with ~$M_2\approx |\mu|$ ~and ~$\mu<0$ ~where ~$R_b$ ~
remains large, ~$\sim0.2180$, ~even for the lighter chargino as heavy as ~
$90$ GeV. ~It is an interesting mixture of the up-higgsino and gaugino.
The r\^ole of various phenomenological  constraints is disscussed in detail in
analitic form and importance of small but {\it non-negligible} left-right
mixing in the stop sector is emphasized in this context.

The large ~$\tan\beta$ ~option for enhancement of ~$R_b$ ~is also summarized,
with similar conclusions to those presented in the earlier papers. The
available data do not rule out the possibility of large ~$R_b$ ~in the large ~
$\tan\beta$ ~case.

We conclude that the new LEP1.5 limit, ~$m_{C_1}>65$ GeV ~still leaves 
open the possibility of a supersymmetric explanation of ~$R_b$ ~up 
to ~$0.2180$. ~We also conclude that  ~$R_b>0.2175$ ~even for ~
$m_{C_1} = M_{\tilde t_2} = 90$ GeV ~both in small ~$\tan\beta$ ~
and  large ~$\tan\beta$ ~cases. Thus LEP2 may not resolve this question.

Finally, we stress that a good quality of the global fit requires the 
hierarchy ~$M_{\tilde t_2}\ll M_{\tilde t_1}$ ~(i.e. ~
$M_{\tilde t_R}\ll M_{\tilde t_L}$). ~This hierarchy is natural if the 
low energy values of the soft squark
masses have their origin in the renormalization group evolution from 
the GUT scale with the initial condition ~$m^2_0\gg M^2_2$.

\newpage
\noindent {\bf Acknowledgments:} We are grateful to M. Olechowski for
an independent check of our numerical code for ~$\delta R_b^{SUSY}$ ~
and to W. de Boer for communicating to us his unpublished results
which agree wuth ours.
We would also like to thank R. Hempfling for discussion.
\noindent P.H. Ch. would like to thank Max--Planck--Institut 
f\"ur Physik for warm hospitality during his stay in Munich where 
part of this work was done.

\newpage
{\bf APPENDIX A.}
\vskip 0.3cm

Here we collect formulae for the remaining contributions to the ~
$Z^0\overline bb$ ~vertex.

The contribution from charged Higgs boson is given by:
\begin{eqnarray}
\delta F^{(t)}_V&=&{e^3\over8s^3_Wc^3_W}
\left[\left(1-{4\over3}s^2_W\right)X_b - {4\over3}s^2_WX_t\right] ~
f_{ffs}(M^2_Z;m_t,M_{H^\pm},m_t)\nonumber\\
&-&{e^3\over8s^3_Wc^3_W}
\left[\left(1-{4\over3}s^2_W\right)X_t - {4\over3}s^2_WX_b\right] ~
m_t^2 ~c_0(m_t,M_{H^\pm},m_t)
\end{eqnarray}

\begin{eqnarray}
\delta F^{(t)}_A&=&-{e^3\over8s^3_Wc^3_W}
\left[\left(1-{4\over3}s^2_W\right)X_b + {4\over3}s^2_WX_t\right] ~
f_{ffs}(M^2_Z;m_t,M_{H^\pm},m_t)\nonumber\\
&-&{e^3\over8s^3_Wc^3_W}
\left[\left(1-{4\over3}s^2_W\right)X_t + {4\over3}s^2_WX_b\right] ~
m^2_t ~c_0(m_t,M_{H^\pm},m_t)
\end{eqnarray}
\begin{eqnarray}
\delta F^{(H^{\pm})}_V=-{e^3\over8s^3_Wc^3_W}\left(1-2s^2_W\right)
\left(X_b+X_t\right) ~f_{ssf}(M^2_Z;M_{H^\pm},m_t,M_{H^\pm})
\end{eqnarray}
\begin{eqnarray}
\delta F^{(H^{\pm})}_A={e^3\over8s^3_Wc^3_W}\left(1-2s^2_W\right)
\left(X_b-X_t\right) ~f_{ssf}(M_Z^2;M_{H^\pm},m_t,M_{H^\pm})
\end{eqnarray}
where we used the abbreviations ~
\begin{eqnarray}
X_b\equiv\left({m_b\over M_Z}\tan\beta\right)^2, ~~~~
X_t\equiv\left({m_t\over M_Z}\cot\beta\right)^2
\end{eqnarray}
As is well known,  contribution of the charged Higgs boson 
to ~$R_b$ ~is negative \cite{BF}.
Positive contribution to ~$R_b$ ~is provided by the neutral Higgs bosons. 
\begin{eqnarray}
\delta F^{(b,A^0)}_V =-{e^3\over16s^3_Wc^3_W}\left(1-{4\over3}s^2_W\right) ~
X_b ~f_{ffs}(M^2_Z;m_b,M_{A^0},m_b)
\end{eqnarray}
\begin{eqnarray}
\delta F^{(b,A^0)}_A = {e^3\over16s^3_Wc^3_W} ~X_b ~
f_{ffs}(M^2_Z;m_b,M_{A^0},m_b)
\end{eqnarray}

\begin{eqnarray}
\delta F^{(b,H^0,h^0)}_V=-{e^3\over16s^3_Wc^3_W}
\left(1-{4\over3}s^2_W\right) ~X_b 
\left[\cos^2\alpha ~f_{ffs}(M^2_Z;m_b,M_{H^0},m_b) 
\right.\nonumber\\
+\left.\sin^2\alpha ~f_{ffs}(M^2_Z;m_b,M_{h^0},m_b)\right]\phantom{aaa}
\end{eqnarray}
\begin{eqnarray}
\delta F^{(b,H^0,h^0)}_A={e^3\over16s^3_Wc^3_W} ~X_b
\left[\cos^2\alpha ~f_{ffs}(M^2_Z;m_b,M_{H^0},m_b) 
\right.\nonumber\\
+\left.\sin^2\alpha ~f_{ffs}(M^2_Z;m_b,M_{h^0},m_b)\right]\phantom{aaa}
\end{eqnarray}

\begin{eqnarray}
\delta F^{(b,A^0,H^0,h^0)}_V = 0
\end{eqnarray}
\begin{eqnarray}
\delta F^{(b,A^0,H^0,h^0)}_A=-{e^3\over4s^3_Wc^3_W} ~X_b 
\left[\cos^2\alpha ~f_{ssf}(M^2_Z;M_{A^0},m_b,M_{H^0}) 
\right.\nonumber\\
+\left.\sin^2\alpha ~f_{ssf}(M^2_Z;M_{A^0},m_b,M_{h^0})\right]\phantom{aaa}
\end{eqnarray}

where ~$\alpha$ ~is the neutral ~$CP-$even Higgs bosons mixing angle 
\cite{ERZ}.

For completeness we display also formulae for the neutralino - sbottom 
contribution.
\begin{eqnarray}
\delta F^{(\tilde b)}_V=
-{e\over4s_Wc_W}\sum_{n,m,l}V^{l,m}_{Z\tilde b\tilde b}
\left(L^{m,n*}_{b\tilde bN}L^{l,n}_{b\tilde bN}
     +R^{m,n*}_{b\tilde bN}R^{l,n}_{b\tilde bN}\right)\nonumber\\
\times f_{ssf}(M^2_Z;M_{\tilde b_m},m_{N_n},M_{\tilde b_l})\phantom{aaa}
\end{eqnarray}
\begin{eqnarray}
\delta F^{(\tilde b)}_A=
-{e\over4s_Wc_W}\sum_{n,m,l}V^{l,m}_{Z\tilde b\tilde b}
\left(L^{m,n*}_{b\tilde bN}L^{l,n}_{b\tilde bN}
     -R^{m,n*}_{b\tilde bN}R^{l,n}_{b\tilde bN}\right)\nonumber\\
\times f_{ssf}(M_Z^2;M_{\tilde b_m},m_{N_n},M_{\tilde b_l})\phantom{aaa}
\end{eqnarray}

\begin{eqnarray}
\delta F^{(N)}_V=
-{e\over2s_Wc_W}\sum_{n,m,l}
\left(R^{l,m}_{ZNN}L^{n,l*}_{b\tilde bN}L^{n,m}_{b\tilde bN}
     +L^{l,m}_{ZNN}R^{n,l*}_{b\tilde bN}R^{n,m}_{b\tilde bN}\right)\nonumber\\
\times f_{ffs}(M^2_Z;m_{N_m},M_{\tilde b_m},m_{N_l})\phantom{aaaa}
\nonumber\\
+{e\over2s_Wc_W}\sum_{n,m,l}
\left(L^{l,m}_{ZNN}L^{n,l*}_{b\tilde bN}L^{n,m}_{b\tilde bN}
     +R^{l,m}_{ZNN}R^{n,l*}_{b\tilde bN}R^{n,m}_{b\tilde bN}\right)\nonumber\\
\times m_{N_m}m_{N_l}c_0(M^2_Z;m_{N_m},M_{\tilde b_m},m_{N_l})\phantom{aaa}
\end{eqnarray}
\begin{eqnarray}
\delta F^{(N)}_A=
-{e\over2s_Wc_W}\sum_{n,m,l}
\left(R^{l,m}_{ZNN}L^{n,l*}_{b\tilde bN}L^{n,m}_{b\tilde bN}
     -L^{l,m}_{ZNN}R^{n,l*}_{b\tilde bN}R^{n,m}_{b\tilde bN}\right)\nonumber\\
\times f_{ffs}(M^2_Z;m_{N_m},M_{\tilde b_m},m_{N_l})\phantom{aaaa}
\nonumber\\
+{e\over2s_Wc_W}\sum_{n,m,l}
\left(L^{l,m}_{ZNN}L^{n,l*}_{b\tilde bN}L^{n,m}_{b\tilde bN}
     -R^{l,m}_{ZNN}R^{n,l*}_{b\tilde bN}R^{n,m}_{b\tilde bN}\right)\nonumber\\
\times m_{N_m}m_{N_l}c_0(M^2_Z;m_{N_m},M_{\tilde b_m},m_{N_l})\phantom{aaa}
\end{eqnarray}
where the couplings can be found in \cite{ROS} and read:
\begin{eqnarray}
L^{l,m}_{ZNN}&=&Z_N^{4l*}Z_N^{4m}-Z_N^{3l*}Z_N^{3m} \nonumber\\
R^{l,m}_{ZNN}&=&Z_N^{3l}Z_N^{3m*}-Z_N^{4l}Z_N^{4m*}
\end{eqnarray}
\begin{eqnarray}
V_{Z\tilde b\tilde b}^{l,m} = B^{1l}B^{1m} - {2\over3}s^2_W\delta^{lm}
\end{eqnarray}
\begin{eqnarray}
L^{n,l}_{b\tilde bN}&=&{e\over\sqrt2s_Wc_W}\left[B^{1n}({1\over3}s_WZ_N^{1l}
-c_WZ^{2l}_N) + {m_b\over M_Z\cos\beta}B^{2n}Z_N^{3l}\right]\nonumber\\
R^{n,l}_{b\tilde bN}&=&{\sqrt2e\over3c_W}B^{2n}Z_N^{1l*}
+ {e\over\sqrt2s_Wc_W}{m_b\over M_Z\cos\beta}B^{1n}Z_N^{3l*}
\end{eqnarray}
where the matrix ~$Z_N$ ~diagonalizes the neutralino mass matrix and the
matrix ~$B^{ij}$ ~for sbottom quarks is defined in the same way as the 
matrix ~$T^{ij}$ ~for stops in eq. (\ref{eqn:tij}).
\vskip 0.5cm

\newpage
{\bf APPENDIX B.}
\vskip 0.3cm
Here we give expressions for the functions ~$f_{ssf}$ ~and ~$f_{fss}$
needed to compute SUSY contributions to the ~$Z^0\overline bb$ ~vertex.
\begin{eqnarray}
f_{ssf}(s,m_1,m_2,m_3) &=&
-{1\over2} +{1\over2s}
\left(a_0(m_1)+a_0(m_3)-2a_0(m_2)\right)
\nonumber\\
&+&{s-m^2_1-m_3^2+2m^2_2\over2s}~
b_0(s,m_1,m_3)\\
&+&\left(m^2_2
+ {(m^2_1-m^2_2)(m^2_3-m^2_2)\over s}\right)
c_0(m_1,m_2,m_3)
\nonumber\\
&-&{1\over4}b_0(0,m_2,m_1)
-{1\over4}(m^2_2-m^2_1)~b^{\prime}_0(m_2,m_1)
\nonumber\\
&-&{1\over4}b_0(0,m_2,m_3)
-{1\over4}(m^2_2-m^2_3)~b^{\prime}_0(m_2,m_3)
\nonumber
\label{eqn:f_ssf}
\end{eqnarray}

\begin{eqnarray}
f_{ffs}(s;m_1,m_2,m_3)&=&
{1\over2} + {s+m^2_1+m^2_3-2m^2_2\over2s}~
b_0(s,m_1,m_3) 
\nonumber\\
&-&{1\over2s}\left(a_0(m_1) + a_0(m_3) -2 a_0(m_2)\right)
\nonumber\\
&-&{(m^2_1-m^2_2)(m^2_3-m^2_2)\over s} ~
c_0(m_1,m_2,m_3)\\
&-&{1\over4}b_0(0,m_1,m_2) 
- {1\over4}(m^2_1-m^2_2)~
b^{\prime}_0(m_1,m_2)
\nonumber\\
&-&{1\over4}b_0(0,m_3,m_2) 
- {1\over4}(m^2_3-m^2_2)~
b^{\prime}_0(m_3,m_2)
\nonumber\label{eqn:f_ffs}
\end{eqnarray}

Standard two-point functions used in eqs. (\ref{eqn:f_ssf},\ref{eqn:f_ffs})
read:
\begin{eqnarray}
a_0(m) = m^2\left(\eta - 1 + \log{m^2\over Q^2}\right)
\end{eqnarray}
\begin{eqnarray}
b_0(s,m_1,m_2) =\eta + \int_0^1dx\log{x(x-1)s + xm_1^2 + (1-x)m^2_2\over Q^2}
\end{eqnarray}
$Q^2$ is the ~${\overline {MS}}$ ~renormalization scale and ~
$\eta\equiv 2/(d-4)$. ~Derivative at ~$s=0$ ~of ~$b_0(s,m_1,m_2)$ ~reads:
\begin{eqnarray}
b_0^{\prime}(m_1,m_2)=-{1\over2}{m_1^2+m_2^2\over(m_1^2-m_2^2)^2}
+ {m_1^2m_2^2\over(m_1^2-m_2^2)^3}\log{m_1^2\over m_2^2}
\end{eqnarray}
Finally, the ~$c_0(m1,m2,m3)$ ~function is defined as:
\begin{eqnarray}
c_0(m1,m2,m3) = \int {d^4k\over\pi^2} 
{i\over [k^2-m_1^2][(k+p)^2 -m_2^2][(k+p+q)^2 -m_3^2]}
\end{eqnarray}
In the case of the ~$Z^0\overline bb$ ~vertex ~$(p+q)^2=M^2_Z$ ~and 
we ~$p^2=q^2=m_b^2\approx0$. ~

\newpage
\begin{figwindow}[0,l,%
{\epsfig{bbllx=40pt,bblly=175pt,bburx=550pt,bbury=670pt,figure=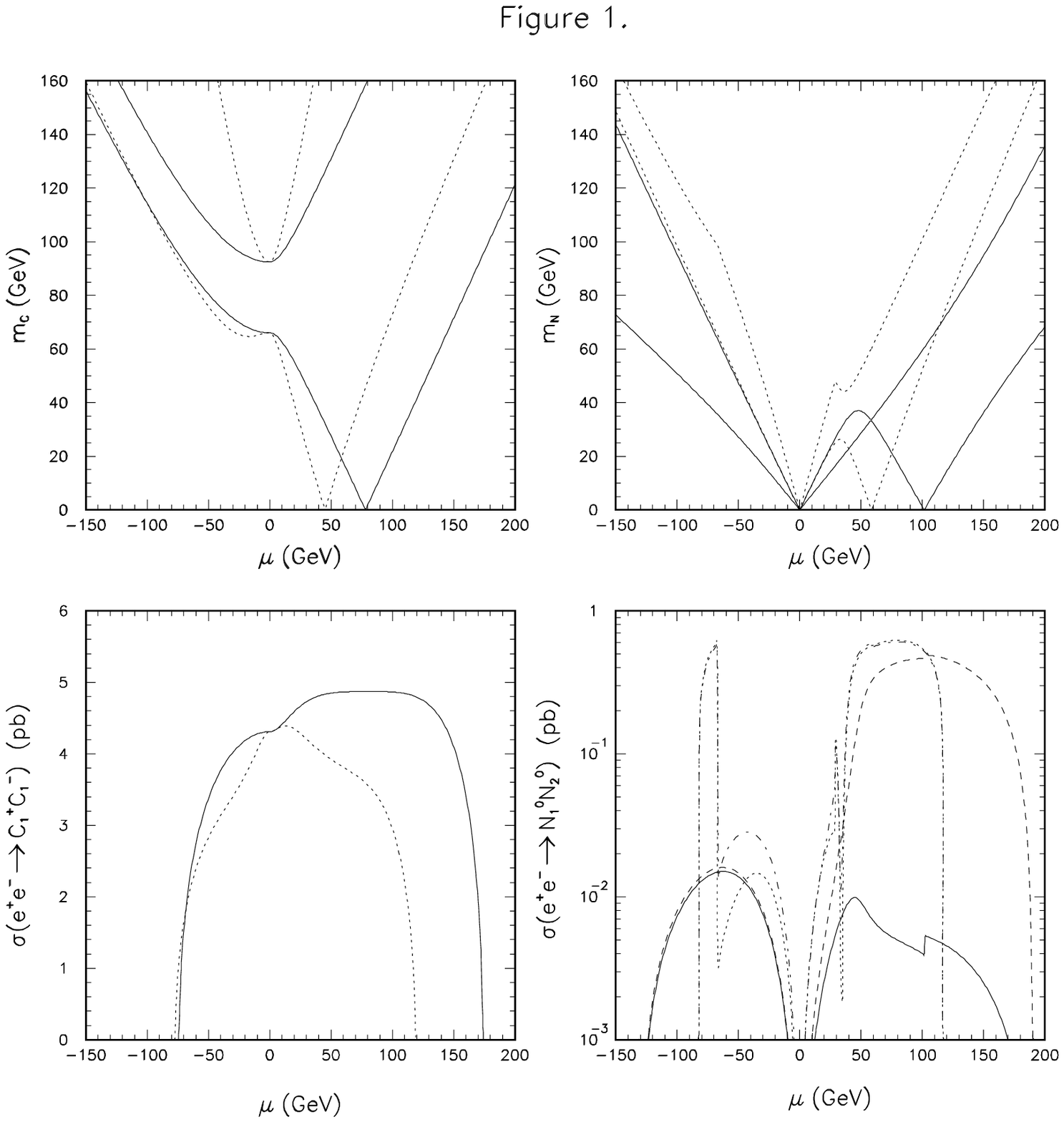,width=\textwidth,clip=}},%
{\protect Masses of the two charginos and two lightest neutralinos
as a function of ~$\mu$ ~for ~$r=1$ ~(solid lines) ~and ~$r=3$ ~(dotted lines)
for ~$\tan\beta=1.4$. ~
In the lower pannels the
corresponding production cross sections of ~
$C_1^+C_1^-$ ~and ~$N_1^0N_2^0$ ~at LEP2 ~($s^{1/2}=190$ GeV) ~are shown.
Masses of the sparticles exchanged in the ~$t$ ~channel are taken to be ~
$M_{\tilde\nu_e}=M_{\tilde e_R}=500$ GeV. ~
For neutralinos also the case  ~$M_{\tilde e_R}=50$ GeV ~(dashed and 
dash-dotted lines for ~$r=1$ ~and ~$r=3$ ~respectively) is shown.}]
\end{figwindow} 

\newpage
\begin{figwindow}[0,l,%
{\epsfig{bbllx=130pt,bblly=260pt,bburx=580pt,bbury=595pt,figure=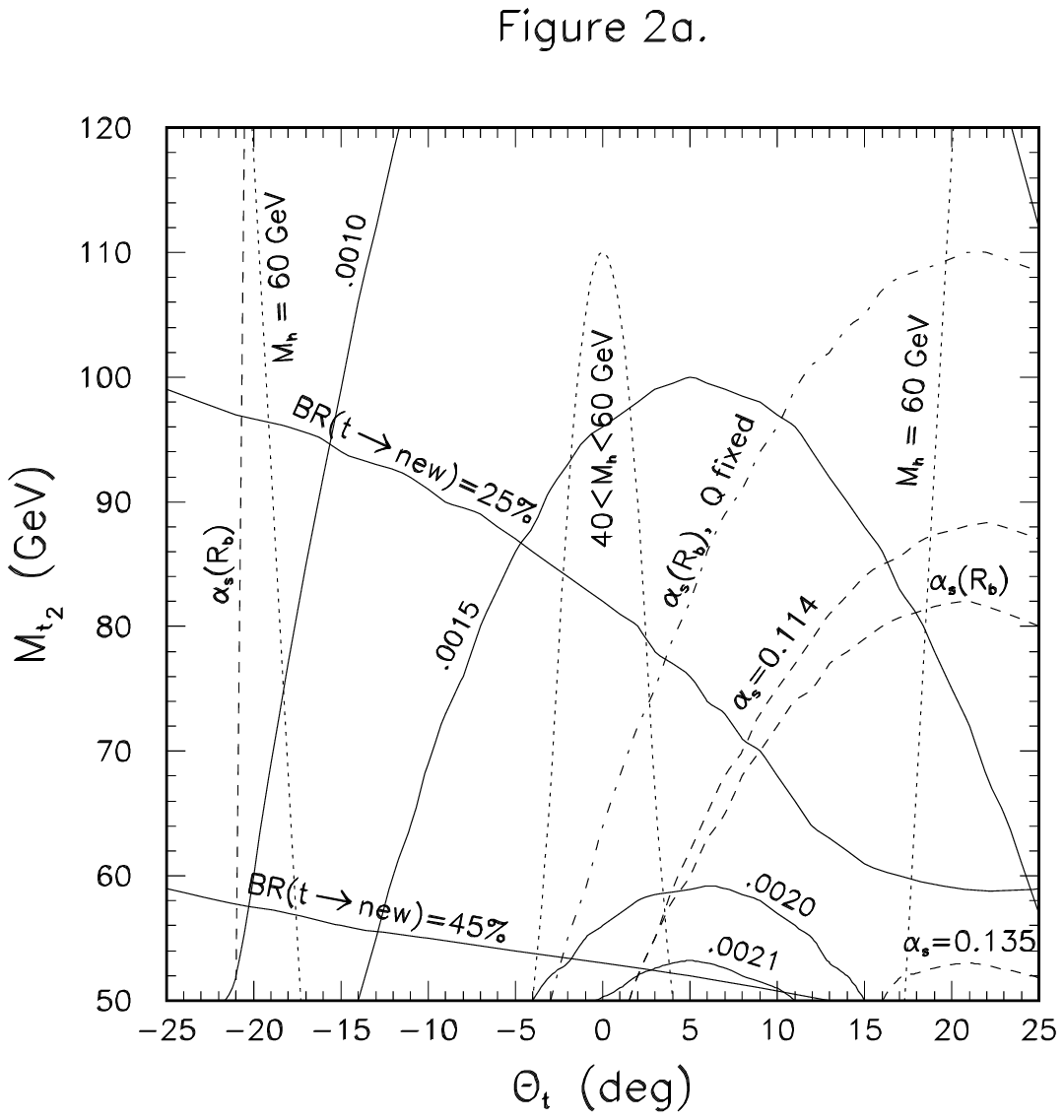,width=\textwidth,clip=}},%
{\protect Contours of constant ~$\delta R_b^{SUSY}$ ~
(solid lines) and various constraints in the plane ~
$(\theta_t, ~M_{\tilde t_2})$ ~for ~$m_t=170$ GeV, ~
$\tan\beta=1.4$, ~$M_2 = -\mu = 58$ GeV ~($m_{C_1}=85$ GeV),  ~
$M_A=M_{\tilde t_1}=1$ TeV. ~
Dashed and dash-doted lines show the ~
$b\rightarrow s\gamma$ ~constraint with different treatement of ~$\alpha_s$: ~
$\alpha_s(R_b)$ ~denotes the curves obtained with ~
$\alpha_s(M_Z)=0.123-4\delta R_b^{SUSY}$ ~and with the renormalization 
scale ~$Q$ ~varied in the range ~$(m_b/2, ~2m_b)$. ~The curve 
with  ~$Q$ ~fixed correspond to ~$Q=m_b=4.7$ GeV ~and ~$\alpha_s(M_Z)=0.123$ ~
(for more details see the text). Dotted lines illustrate the Higgs boson 
mass constraint.
The allowed region is bounded from below by the ~$BR(t\rightarrow new)=45$ \% ~
curve and the parabolic ~$b\rightarrow s\gamma$ ~curve ~
$\alpha_s(R_b)$ ~and from the left- and right- hand sides by the dotted
curves ~$M_h=60$ GeV. ~The area below 
the central dotted curve is also excluded.}]
\end{figwindow} 

\newpage
\begin{figwindow}[0,l,%
{\epsfig{bbllx=130pt,bblly=260pt,bburx=580pt,bbury=595pt,figure=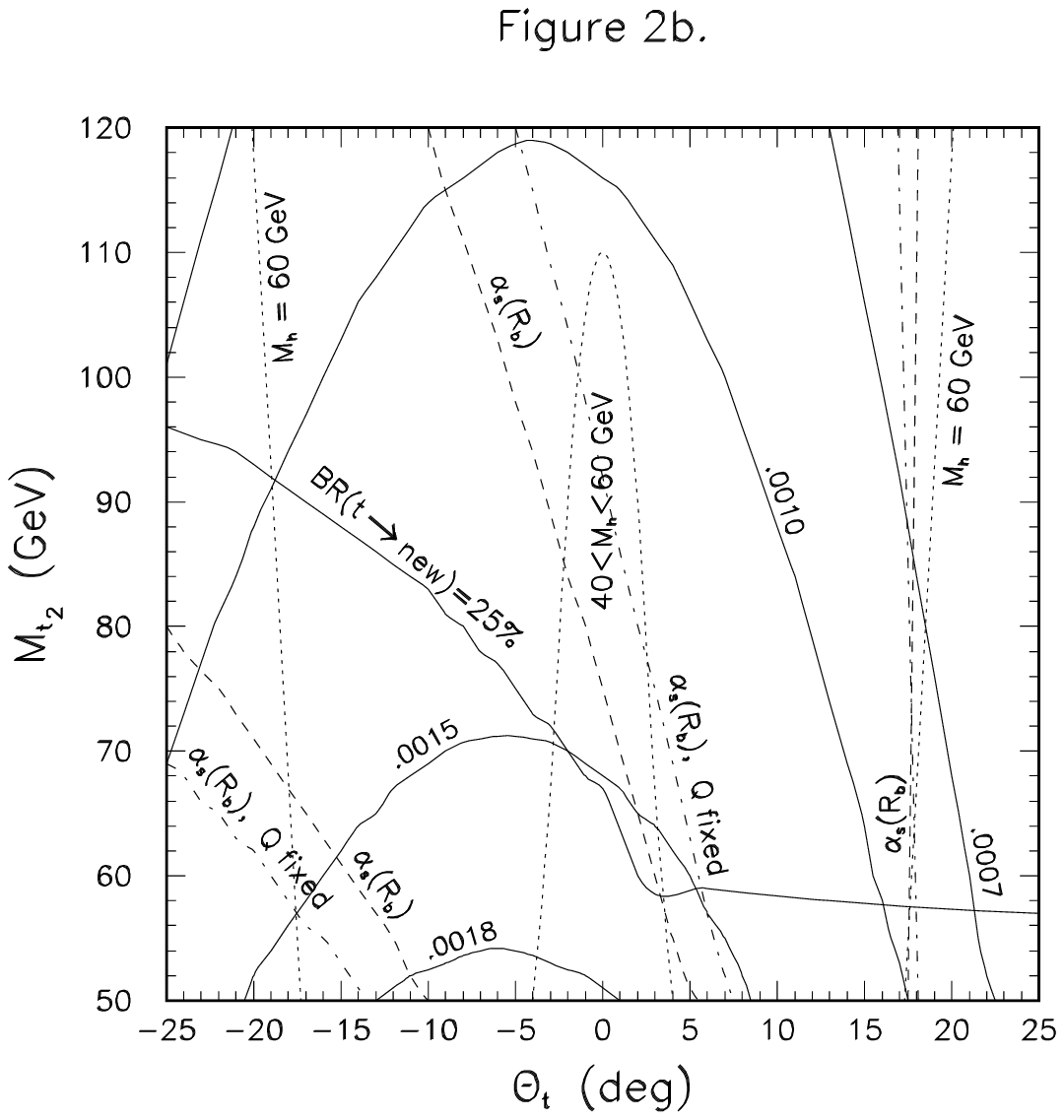,width=\textwidth,clip=}},%
{\protect As in Fig. 2 but for ~$\mu>0$ ~and ~$r=1.5$. ~
$\mu=85.5$ GeV  ~($m_{C_1}=55$ GeV),  ~$M_A=M_{\tilde t_1}=1$ TeV. ~
The allowed region is for ~$\theta_t > 0$, ~between the two dashed curves 
denoted by ~$\alpha_s(R_b)$ ~(with exclusion of the central area ~
$40 <M_h < 60$ GeV) ~and for ~$\theta_t < 0$, ~below the dashed curve ~
$\alpha_s(R_b)$ ~ (bounded from the left by the dotted curve ~$M_h=60$ GeV).}]
\end{figwindow} 

\newpage
\begin{figwindow}[0,l,%
{\epsfig{bbllx=50pt,bblly=165pt,bburx=555pt,bbury=684pt,figure=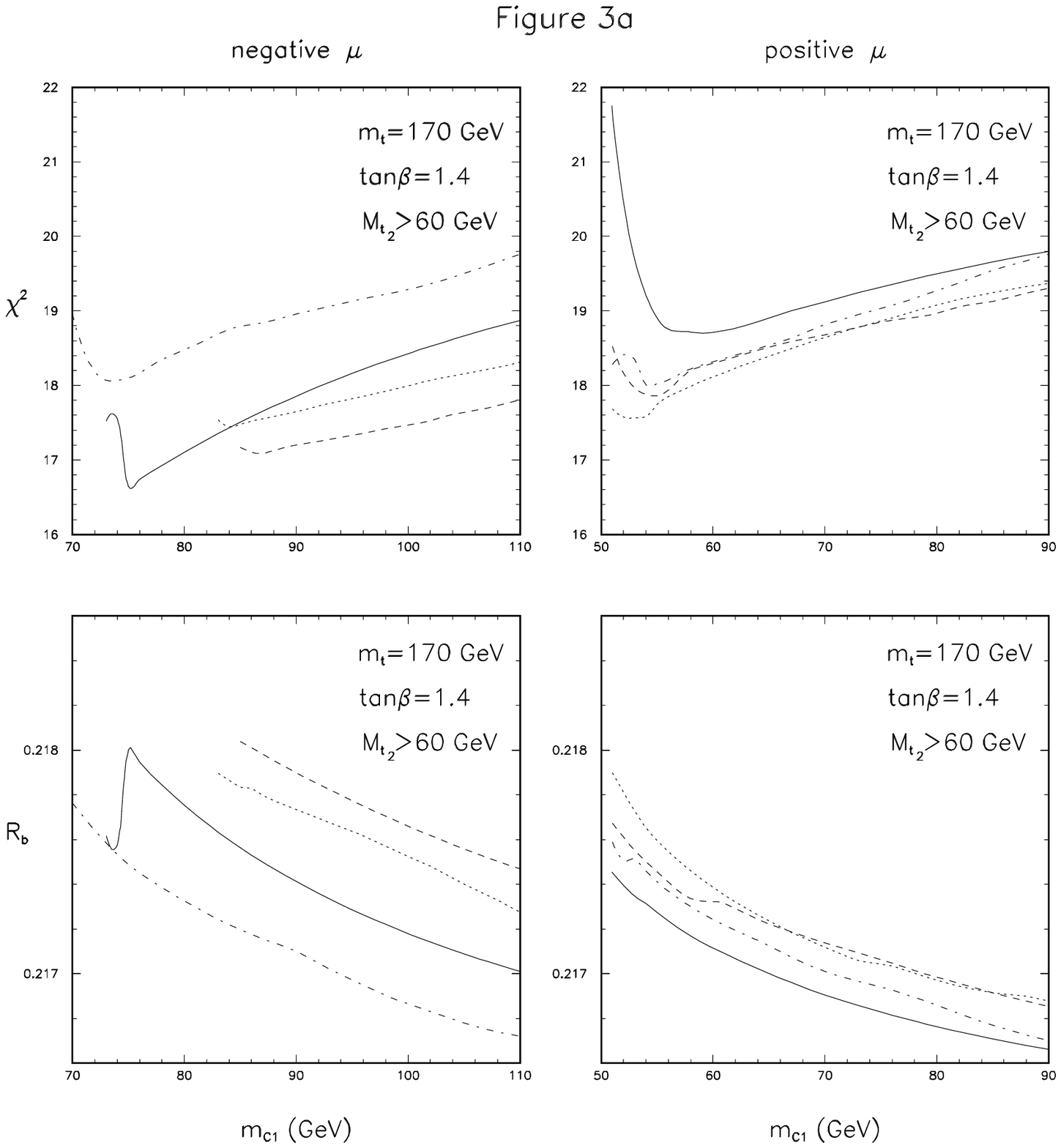,width=\textwidth,clip=}},%
{\protect $\chi^2$ as a function of ~$m_{C_1}$ ~for ~
$r\equiv M_2/|\mu|=0.5$ ~(solid lines), ~$1$ ~(dashed), ~$1.5$ ~(dotted), 
and ~$3$ ~(dash-dotted) for both signs of ~$\mu$ ~for ~$m_t=170$ GeV, ~
$\tan\beta=1.4$, ~$M_A=M_{\tilde t_1}=1$ TeV. ~In lower pannels the best 
values of ~$R_b$ ~with the restriction ~$\chi^2< \chi^2_{min} +1$ ~(here ~
$\chi^2_{min}$ ~denotes the best ~$\chi^2$ ~for fixed value of ~$m_{C_1}$) ~
are shown. In addition we required ~$M_{\tilde t_2}>60$ GeV.}]
\end{figwindow} 

\newpage
\begin{figwindow}[0,l,%
{\epsfig{bbllx=50pt,bblly=165pt,bburx=555pt,bbury=684pt,figure=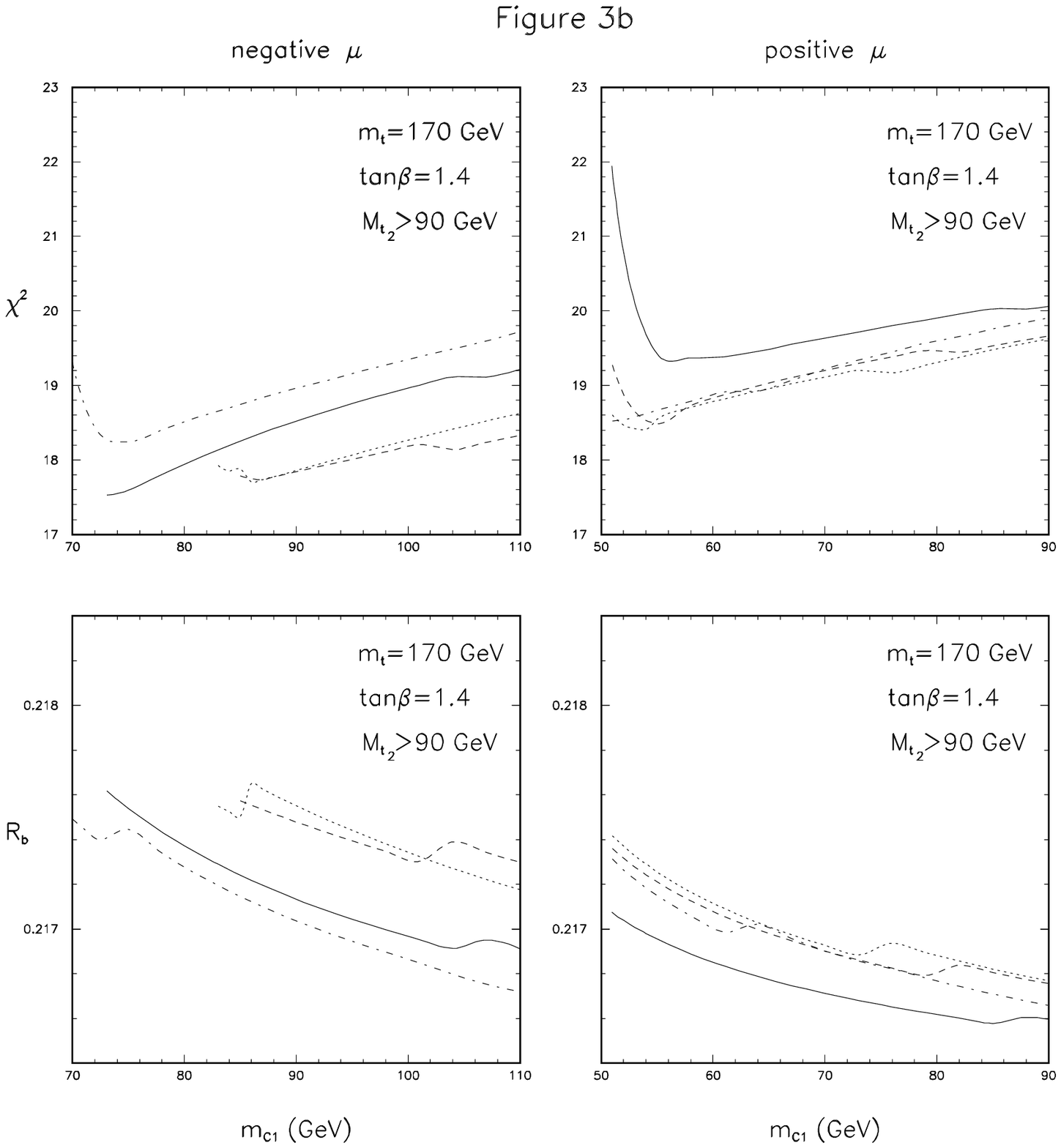,width=\textwidth,clip=}},%
{\protect As in Fig. 4 but with condition ~
$M_{\tilde t_2}>90$ GeV.}]
\end{figwindow} 

\newpage
\begin{figwindow}[0,l,%
{\epsfig{bbllx=50pt,bblly=165pt,bburx=555pt,bbury=684pt,figure=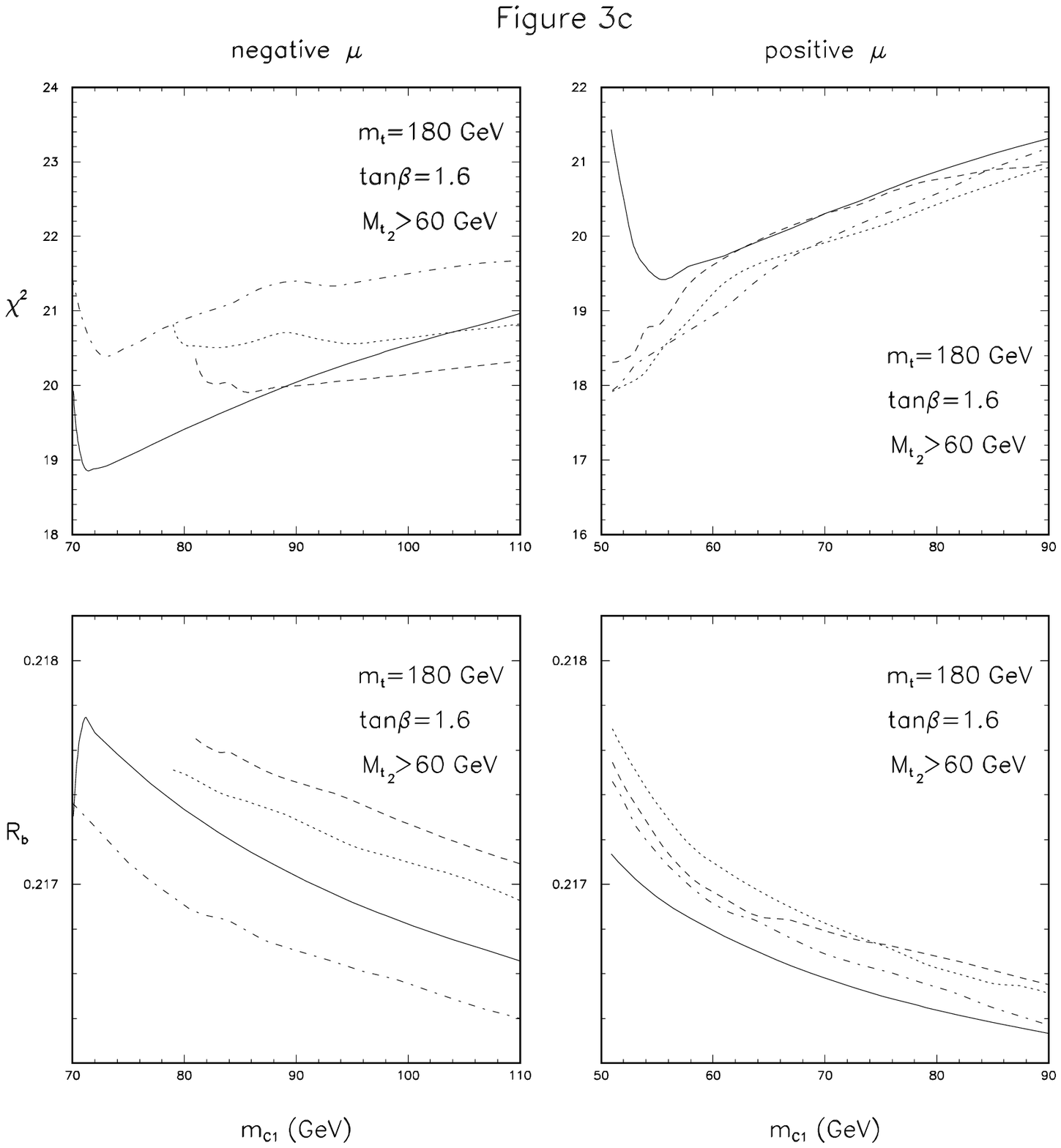,width=\textwidth,clip=}},%
{\protect As in Fig. 4 but for ~$m_t=180$.}]
\end{figwindow} 

\newpage
\begin{figwindow}[0,l,%
{\epsfig{bbllx=130pt,bblly=260pt,bburx=580pt,bbury=595pt,figure=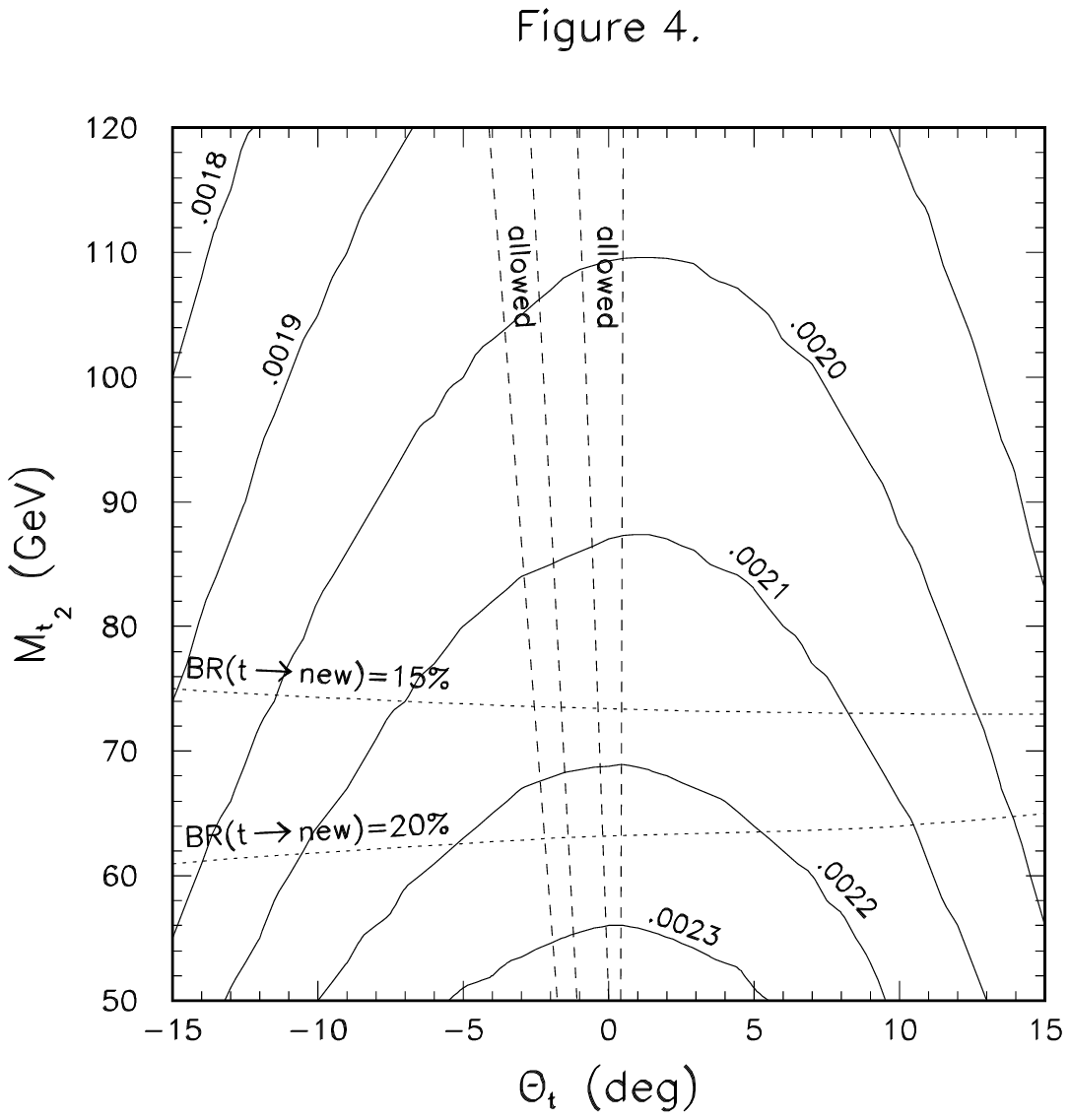,width=\textwidth,clip=}},%
{\protect Contours of constant ~$\delta R_b^{SUSY}$ ~
(solid lines) and allowed by ~$BR(b\rightarrow s\gamma)$ ~regions in the ~
$(\theta_t, ~M_{\tilde t_2})$ ~plane for ~$m_t=170$ GeV, ~
$\tan\beta=50$, ~$M_2 = 1.5\mu$, ~$m_{C_1}=65$ GeV,  ~
$M_{\tilde t_1}=1$ TeV, ~$M_A=55$ GeV ~and ~$M_{\tilde b_R}=130$ GeV. ~
Contours of constant ~$\sum_iBR(t\rightarrow\tilde t_2N^0_i)$ ~are also shown
(dotted lines).}]
\end{figwindow} 

\newpage
\begin{figwindow}[0,l,%
{\epsfig{bbllx=50pt,bblly=165pt,bburx=555pt,bbury=678pt,figure=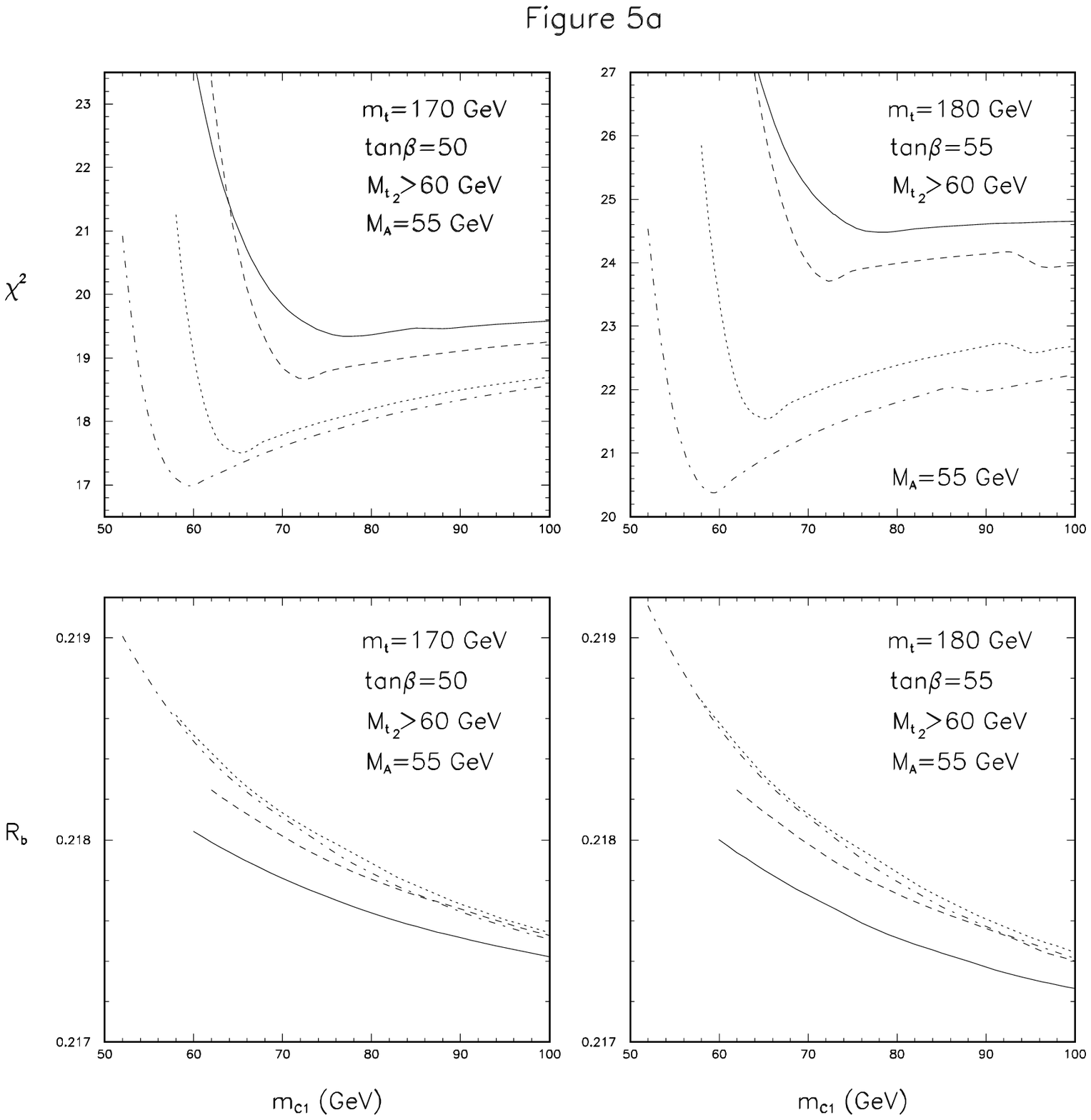,width=\textwidth,clip=}},%
{\protect $\chi^2$ as a function of ~$m_{C_1}$ ~($\mu>0$) ~
for ~$r\equiv M_2/|\mu|=1$ ~(solid lines), ~$1.5$ ~(dashed), ~$3$ ~(dotted), 
and ~$5$ ~(dash-dotted) for ~$m_t=170$ GeV, ~$\tan\beta=50$ ~and ~180 GeV, ~
$\tan\beta=55$. ~$M_{\tilde t_1}=1$ TeV, ~$M_A=55$ GeV, ~
$M_{\tilde b_R}=130$ GeV.  ~In lower pannels the best 
values of ~$R_b$ ~with the restriction ~$\chi^2< \chi^2_{min} +1$ ~(here ~
$\chi^2_{min}$ ~denotes the best ~$\chi^2$ ~for fixed value of ~$m_{C_1}$) ~
are shown. In addition we required ~$M_{\tilde t_2}>60$ GeV.}]
\end{figwindow} 

\newpage
\begin{figwindow}[0,l,%
{\epsfig{bbllx=50pt,bblly=165pt,bburx=555pt,bbury=678pt,figure=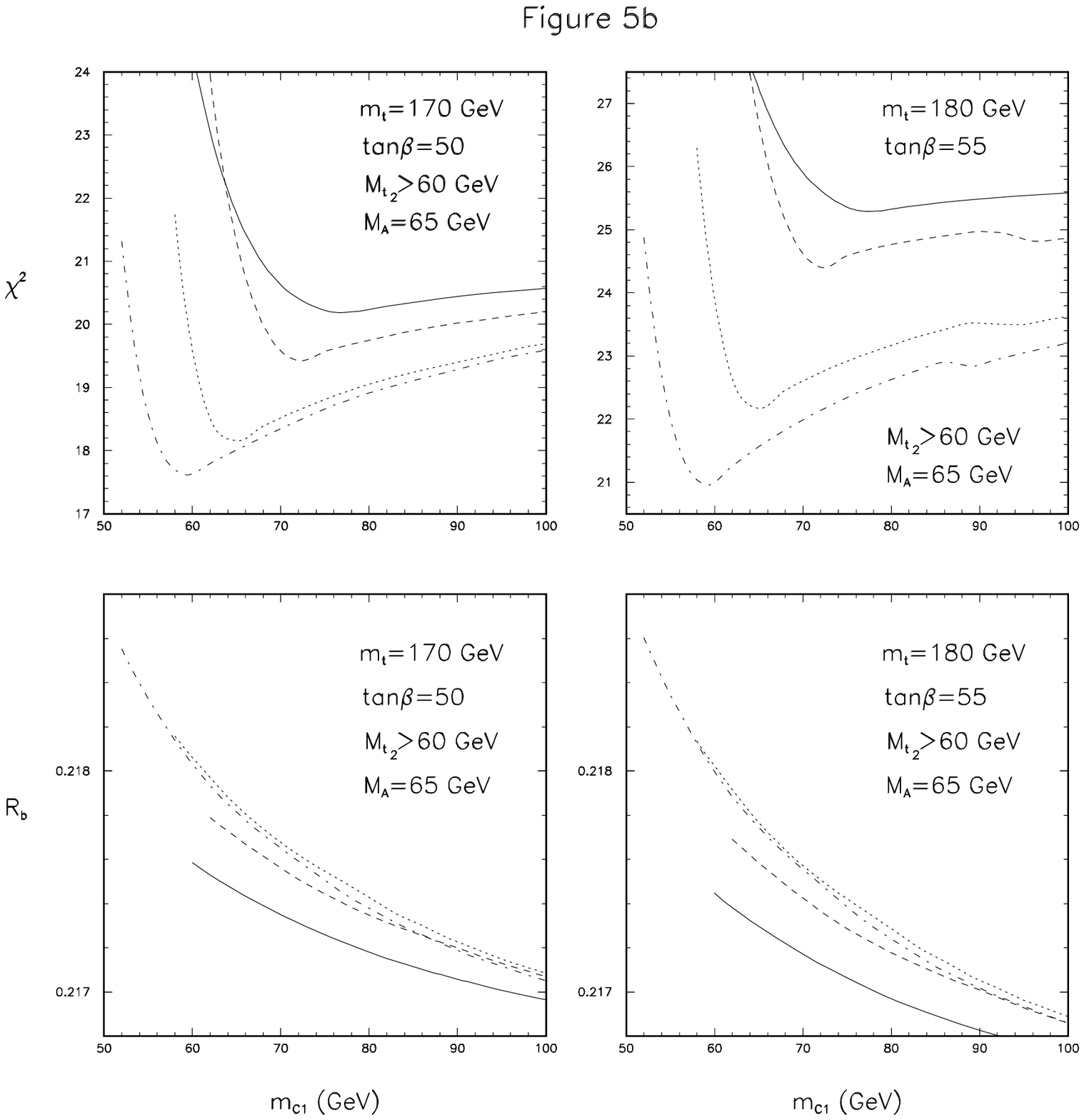,width=\textwidth,clip=}},%
{\protect As in Fig. 8 but for ~$M_A=65$ GeV.}]
\end{figwindow} 

\end{document}